\documentclass{elsarticle}

\usepackage[linesnumbered,noend,vlined, scleft,nofillcomment,ruled]{algorithm2e}

\usepackage{amsthm,graphicx}
\usepackage{lineno,hyperref}
\usepackage{xspace}
\usepackage{xcolor}
\modulolinenumbers[5]
\usepackage{multirow}
\usepackage{color}

\newtheorem{theorem}{Theorem}

\newtheorem{corollary}[theorem]{Corollary}
\newtheorem{fact}[theorem]{Fact}
\newtheorem{definition}{Definition}

\newcommand{\Oh}[1]
  {\ensuremath{\mathcal{O}\!\left( {#1} \right)}}

\newcommand {\changed}[1]{#1}

\newcommand{\DAC}{\ensuremath{\mathrm{DAC}}}
\newcommand{\kt}{$k^2$-tree\xspace}
\newcommand{\ktree}{\kt}
\newcommand{\koct}{$k^3$-tree\xspace}
\newcommand{\koctdac}{$k^3$-tree$^{\DAC}$\xspace}
\newcommand{\ktplain}{$k^2$-tree$^p$\xspace}
\newcommand{\ktdac}{$k^2$-tree$^{\DAC}$\xspace}
\newcommand{\dblp}{\textsf{dblp}\xspace}
\newcommand{\enwiki}{\textsf{enwiki}\xspace}
\newcommand{\indo}{\textsf{indochina}\xspace}
\newcommand{\uk}{\textsf{uk}\xspace}
\newcommand{\giss}{\textsf{GIS-sparse}\xspace}
\newcommand{\gism}{\textsf{GIS-med}\xspace}
\newcommand{\gisd}{\textsf{GIS-dense}\xspace}
\newcommand{\rdfs}{\textsf{RDF-sparse}\xspace}
\newcommand{\rdfm}{\textsf{RDF-med}\xspace}
\newcommand{\rdfd}{\textsf{RDF-dense}\xspace}
\newcommand{\mdta}{\textsf{mdt500}\xspace}
\newcommand{\mdtb}{\textsf{mdt700}\xspace}
\newcommand{\mdtc}{\textsf{mdtmed}\xspace}
\newcommand{\LCP}{\ensuremath{\mathrm{LCP}}}

\newcommand{\hpqt}{\textsf{hpqt}\xspace}
\newcommand{\hpqtp}{\textsf{hpqt$^p$}\xspace}
\newcommand{\hpqtR}{\textsf{hpqt$^{c}$}\xspace}
\newcommand{\hpqtpdac}{\textsf{hpqt$^{p+\DAC}$}\xspace}
\newcommand{\hpqtRdac}{\textsf{hpqt$^{c+\DAC}$}\xspace}

\newcommand{\pdt}{PDT\xspace}
\newcommand{\pdth}{PDT-hollow\xspace}
\newcommand{\pdtrp}{PDT-RP\xspace}

\newcommand{\no}[1]{}

  {\color{gray}}%
  {}
  
\newenvironment{example}%
  {\paragraph{Example}}%
  {\qed \medskip }

\journal{TBD}

\begin{document}

\begin{frontmatter}

\title{Faster Compressed Quadtrees}

\author{Guillermo de Bernardo$^1$, Travis Gagie$^2$, 
Susana Ladra$^1$,\\ Gonzalo Navarro$^{3,4}$ and Diego Seco$^{3,5}$\\
\ \\
 $^1$ Universidade da Coruña, CITIC, Database Lab, Spain\\
 $^2$ Faculty of Computer Science, Dalhousie University, Canada\\
 $^3$ IMFD --- Millennium Institute for Foundational Research on Data, Chile\\
 $^4$ Department of Computer Science, University of Chile, Chile\\
 $^5$ Department of Computer Science, University of Concepci\'on, Chile}
\date{}
\tnotetext[acks]{An early partial version of this paper appeared in {\em Proc. of the Data Compression Conference 2015}.}

\begin{abstract}
Real-world point sets tend to be clustered, so using a machine word for each point is wasteful. In this paper we first show how a compact representation of quadtrees using $\Oh{1}$ bits per node can break this bound on clustered point sets, while offering efficient range searches. We then describe a new compact quadtree representation based on heavy path decompositions, which supports queries faster than previous compact structures. We present experimental evidence showing that our structure is competitive in practice.
\end{abstract}


\end{frontmatter}

\section{Introduction}
\label{sec:introduction}

Storing and querying two-dimensional points sets is fundamental in computational geometry, geographic information systems, graphics, and many other fields.  Most researchers have aimed at designing data structures whose size, measured in machine words, is linear in the number of points.  That is, data structures are considered small if they store a set of $n$ points on a \(u \times u\) grid in $\Oh{n}$ words of $\Oh{\log u}$ bits each.  Using $\Oh{n \log u}$ bits is within a constant factor of optimality when the points are distributed uniformly at random over the grid, but we can often do better on real-world point sets because they tend to be clustered and, therefore, compressible.

Quadtrees~\cite{Mor66,Sam06} store the point's coordinates in implicit form, along a root-to-leaf path per point. Quadtrees may have \(o (n \log u)\) nodes when the points are clustered, because closer points tend to share a longer part of their path. Still, classic quadtrees are implemented with pointers, which take $\Omega(\log u)$ bits per node, and since they use one node per point at the very least, they require \(\Omega (n \log u)\) bits overall; the same happens if we store the explicit coordinates instead of the paths \cite{Gar82}.  

Recently, various authors \cite{BLN14,VM14,BCBNP20} proposed quadtree representations based on succinct trees, which avoid pointers. 
These structures store the coordinates implicitly using the paths, and those paths use $\Oh{1}$ bits per quadtree node. Therefore, they are able to use $o(n\log u)$ bits of space, while offering the same asymptotic query times as traditional structures when supporting edge-by-edge navigation. Venkat and Mount \cite{VM14} noted, however, that
\begin{quotation}
``A method for compressing paths or moving over multiple edges at once using a succinct structure may speed up the many algorithms that rely on traversal of the quadtree.''
\end{quotation}
Some previous data structures, such as skip-quadtrees~\cite{EGS08} and path-decomposed tries~\cite{GO14}, are evidence that quadtree variants can indeed use $O (1)$ bits per node while moving over multiple edges at once. The authors of skip-quadtrees only aimed at a space bound of $O (n \log u)$ bits and did not give an implementation, while the authors of path-decomposed tries gave a mainly experimental analyses.

This paper contains two main contributions:
\begin{enumerate}
    \item We give a space analysis of quadtree data structures as a function of the amount of clustering of the point set, showing that compressed quadtrees can use $o(n\log u)$ bits of space on clustered points. We also show that quadtree queries speed up on clustered points.
    \item We present the first compressed quadtree data structure that, within that space, uses heavy-path decomposition in order to provide one-step navigation over multiple edges, thereby speeding up queries.
\end{enumerate}

After describing the compressed quadtree data structure in Section~\ref{sec:related},
contribution 1 is provided in Section~\ref{sec:space}, and contribution 2 in Sections~\ref{sec:structure} (which describes the new structure) and Section~\ref{sec:membership} (which gives the new query algorithms).  In Section~\ref{sec:practice} we describe some practical improvements and in Section~\ref{sec:experiments} we show experimentally that our structure is competitive, and in particular that it outperforms current alternatives when retrieving isolated points. We conclude in
Section~\ref{sec:conclusions}.

\section{Basic Concepts}
\label{sec:related}

\subsection{Model of computation}

Like most of the work on compressed data structures, we assume the RAM model of computation, where the machine word holds $\Theta(\log u)$ bits and can perform all the usual arithmetic and bitwise operations operations on words in constant time. Note $\log n = \Oh{\log u}$ because $n \le u^2$.

\subsection{Bitvectors}

A bitvector is an array $B[1,n]$ of bits. We are interested, apart from accessing any bit $B[i]$, in implementing two operations: $rank_b(B,i)$ counts the number of times bit $b$ appears in $B[1,i]$, whereas $select_b(B,j)$ is the position of the $j$th occurrence of bit $b$ in $B$. All these operations can be computed in constant time using only $o(n)$ extra bits on top of $B$ \cite{Cla96,Mun96}.

\subsection{Quadtrees} \label{sec:quadtrees}

There are many kinds of quadtrees. Our definition corresponds to the so-called 
MX-Quadtree \cite{WF90,Sam06}.

\begin{definition}
Let $\cal P$ be a set of $n$ points on a discrete grid $[1,u]^2$. If $n$ is 0, 
then the quadtree for the grid is a leaf storing 0. If $u=1$, then the 
quadtree is a leaf storing 1 if the cell contains a point and 0 if not.
Otherwise, the quadtree is a tree whose root stores a 1 
and has four children, which are the quadtrees of the grid's four quadrants.
We say that a node {\em covers} the area of its subgrid and that it is an 
{\em ancestor} of the points in that subgrid.
\end{definition}

\begin{example}
Figure~\ref{fig:tree} shows an example, taken from Brisaboa et al.~\cite{BCBNP20}.  Notice the order of the quadrants is top-left, top-right, bottom-left, bottom-right, instead of the counterclockwise order customary in mathematics.  This is called the Morton or Z-ordering and it is useful because, assuming $u$ is a power of 2 and the origin is at the top right --- without loss of generality, since we can manipulate the coordinate system to make it so --- the obvious binary encoding of a root-to-leaf path is the interleaving of the binary representations of the corresponding point's $y$- and $x$-coordinates.

\begin{figure}
\begin{center}
\includegraphics[width=\textwidth]{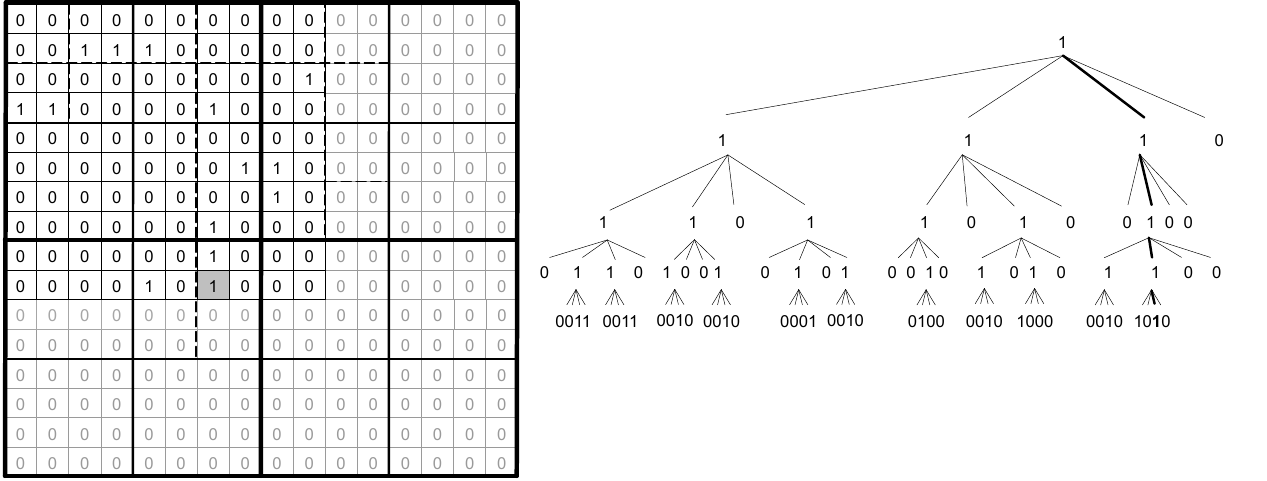}
\caption{A set of points, indicated by 1s, on a \(16 \times 16\) grid (left); the quadtree for those points (right).  The heavy lines in the quadtree indicate the path to the leaf corresponding to the shaded point on the grid.}
\label{fig:tree}
\end{center}
\end{figure}

For example, if we imagine the edges descending from each internal node in Figure~\ref{fig:tree} are labelled \(0, 1, 2, 3\) from left to right, then the thick edges are labelled \(2, 1, 1, 2\); the obvious binary encoding for this path is \(10\,01\,01\,10\).  The coordinates for the shaded point, which corresponds to the leaf at the end of this path, are \((6, 9)\), so interleaving the binary representations 1001 and {\it 0110} of its $y$- and $x$-coordinates also gives \(1\mathit{0}\,0\mathit{1}\,0\mathit{1}\,1\mathit{0}\). \changed{We can interleave a point's coordinates in $\Oh{1}$ time using, for example, pre-computed tables}.
\end{example}

We now summarize a few simple facts on quadtrees.

\begin{fact}
\label{fact:basics}
A quadtree for the set of points $\cal P$ on a grid $[1,u]^2$ has height at most \(\lg u\) and $\Oh{n \log u}$ nodes. A node at depth $j$ covers a square area of size \(2^{\lg(u)-j} \times 2^{\lg(u)-j}\). A leaf storing a 1 is at depth $\lg u$ and hence covers
a single cell; there is exactly one such leaf per point in $\cal P$.
\end{fact}

A quadtree can efficiently find the points lying on a region of the grid.

\begin{definition}
A {\em query} $R \subseteq [1,u]^2$ aims to retrieve the points of $\cal P$
that lie within $R$. The result is denoted ${\cal P} \cap R$. When $R$ is a 
rectangle, the query is called a {\em range query}, and the special
case \(R = [x, x + 1) \times [y, y + 1)\) is called a {\em membership test}
for the point \((x, y)\).  
\end{definition}

Given a query region $R$, the quadtree computes \(\mathcal{P} \cap R\) by starting at the root and visiting all the nodes whose subgrids overlap $R$, reporting the coordinates of every leaf storing 1. In a range query we can determine in constant time whether a node's area overlaps $R$. This reduces the problem of computing the cost of solving a query to that of computing the size of a quadtree.

\begin{fact}
\label{fact:size}
Let $Q = \mathcal{P} \cap R$ be the output of a range query. Then the cost of enumerating $Q$ by traversing all the nodes overlapping $R$ in a quadtree for $\mathcal{P}$ is proportional to the number of nodes in a quadtree for $Q$.
\end{fact}

The Quadtree Complexity Theorem \cite{Kli71,HS79,Sam06} establishes that the number of maximal-area quadtree nodes inside a rectangle $R$ of size $p \times q$, plus their ancestors, is $\Oh{p+q+\log u}$. We traverse all those nodes to 
solve the query $\mathcal{P} \cap R$, plus the paths towards every point inside
$R$. If we pessimistically add $\lg u$ nodes to account for each such path, 
we obtain the following result.

\begin{theorem}
The time complexity for solving the range query $Q = \mathcal{P} \cap R$, 
where $R$ is a rectangle of size $p \times q$, on a quadtree for $\cal P$ over 
a $[1,u]^2$ grid, is $\Oh{p+q+(|Q|+1)\log u}$.
\end{theorem}

If the points in $\cal P$ are clustered, however, then intuitively the root-to-leaf paths in the quadtree will share many nodes and we will use less space and time. The query time can be refined to $\Oh{p+q+\log u+|Q|(1+\log(pq/|Q|))}$ \cite[p.~361]{Nav16}, which shows that the time per reported point decreases on smaller or denser query ranges.
We further exploit this idea to provide more refined space and time bounds on clustered points in the Section~\ref{sec:space}.

\medskip

Quadtrees can be generalized to $d \ge 2$ dimensions, in which case each node 
has $2^d$ children. On a universe $[1,u]^d$, the quadtree still has height 
$\log_{2^d} u^d = \lg u$. The Quadtree Complexity Theorem formula generalizes
to $\Oh{dq^{d-1}+\log u}$ for a hypercube $R$ of side $q$, and consequently
the search time complexity for the query $Q=\mathcal{P} \cap R$ becomes
$\Oh{d q^{d-1} + (|Q|+1) \log u}$.

\subsection{Compressed quadtrees}
\label{sec:comprquad}

Brisaboa, Ladra and Navarro~\cite{BLN14} proposed a compressed quadtree representation called $k^2$-tree (a quadtree corresponds to using $k=2$). It represents a quadtree using exactly $1$ bit per node, by collecting the $0$s and $1$s of the tree in levelwise order (omitting the root). 
They show that, by adding $rank$ and $select$ support to this concatenation of bits, the quadtree can be navigated towards children and parent in constant time: if we identify the node $v$ with the position $i$ so that the $4$ bits describing its children are in $B[4i-3..4i]$ (so the root is $1$), then the identifier of the $j$th child of $v$ is $rank_1(B,4(i-1)+j)+1$, and that of the parent of $v$ is $\lceil select_1(B,i-1)/4\rceil$.

\begin{example}
The quadtree on the right of Figure~\ref{fig:tree} is represented as a bitvector $B$ concatenating the bits $1110$ (the first level), followed by $110110100100$ (the second level), and so on. To traverse the path in bold, we start at the
root node, $i=1$, and take the third child ($j=3$) with 
$i' = rank_1(B,4\cdot 0+3)+1 = 4$. Indeed, the child is the 4th node in a
levelwise traversal of the quadtree. Its second child ($j=2$) is
$i'' = rank_1(B,4\cdot 3+2)+1=10$. Again, the child is the 10th node in the
levelwise traversal. The parent of node $i''$ is $\lceil select_1(B,9)/4\rceil
= 4 = i'$.
\end{example}


There are several other variants of this representation \cite{VM14,BCBNP20}, as well as various techniques to further reduce space.
A major improvement in compression \cite{BLN14} can be obtained in practice by exploiting small-scale regularities that arise in many real-world datasets. To do this, they consider small submatrices of a predefined size (for instance, $4\times4$ or $8\times8$), and only represent the tree up to those submatrices, effectively trimming the lower levels of the tree. The different submatrices that arise are then sorted by frequency and stored explicitly in a matrix vocabulary, and a sequence of matrix identifiers is used as a last level of the tree. Directly-Addressable Codes~\cite{BLN13} (DACs) are used to store and access the sequence.
The $k^2$-tree representation can be naturally extended to $d$ dimensions, hence becoming a 
$k^d$-tree.



\section{Tighter Bounds on Quadtrees of Clustered Point Sets}
\label{sec:space}

We first bound the size of a quadtree when the points can be distributed in $c$ clusters of the same level; then we generalize the result to hierarchical clustering.

\begin{theorem}
\label{thm:space}
Let $\cal P$ be a set of points on the discrete grid $[1,u]^2$.
Let ${\cal P} = {\cal P}_1 \uplus \cdots \uplus {\cal P}_c$ be a partition of
$\cal P$ into $c$ clusters, so that each
${\cal P}_i$ contains $|{\cal P}_i| = n_i$ points lying on a square region of
side $\ell_i$ (the square regions are not necessarily disjoint).
Then the quadtree of $\cal P$ has $\Oh{c \log u + \sum_i n_i \log \ell_i}$ nodes.
\end{theorem}

\begin{proof}
Let $S$ be any \(\ell \times \ell\) square on the grid and
\(N = S \cap \mathcal{P}\) be the points of $\cal P$ that lie within $S$. 
Let $A$ be the set of ancestors of the points in $N$, and
$A'$ be the ancestors of the corners of $S$ (those corners may or may not 
be in $\mathcal{P}$). Since the quadtree of $\cal P$ has maximum height $\lg u$
and $S$ has 4 corners, it holds 
\[|A| \leq |A \cup A'| \leq |A \backslash A'| + |A'| < |A \backslash A'| + 4 \lg u\,.\]

By Fact~\ref{fact:basics},
any ancestor $v$ of a point in $N$ that has depth at most \(\lg (u / \ell)\)
covers all the points in a square of size at least \(2^\ell \times 2^\ell\).  
Therefore, the square must contain at least one corner of $S$, and thus \(v \in A'\).  It follows that
\[|A \backslash A'| \leq |N| (\lg u - \lg (u / \ell)) = |N| \lg \ell\,,\]
so \(|A| < |N| \lg \ell + 4 \lg u\).
Since each cluster ${\cal P}_i$ has $|N|=n_i$ points that lie within a square of
size $\ell_i \times \ell_i$, the result follows.
\end{proof}


\begin{theorem} 
\label{thm:hier}
Let ${\cal P}$ be a set of points on the discrete grid $[1,u]^2$, and
$\mathcal{T}$ be a tree with root $r$. Every node $t \in \cal T$ stores a set 
${\cal P}_t \subseteq \cal P$ of $n_t$ points,
which is the union of the points stored at its children.
The sets ${\cal P}_t$ of all the nodes $t\in\mathcal{T}$ 
at the same depth form a partition of ${\cal P}$ into clusters. The points 
${\cal P}_t$ of every node $t \in \mathcal{T}$ lie on a square region of side 
$\ell_t$; those regions need not be disjoint.
Then the quadtree of $\mathcal{P}$ has $\Oh{\sum_{t \in \mathcal{T} \setminus \{r\}} \log \ell_{p(t)} + \sum_{t \in L} n_t \log \ell_t}$ nodes, where $p(t)$ is the parent of $t$ in $\mathcal{T}$ and $L \subseteq \mathcal{T}$ is the set of leaf nodes in $\mathcal{T}$.
\end{theorem}

\begin{proof}
Applying Theorem~\ref{thm:space} on the non-hierarchical clustering induced by the $c=|L|$ leaves of $\mathcal{T}$, we obtain the upper bound
$\Oh{|L| \log \ell_r + \sum_{t \in L} n_t \log \ell_t}$. We now refine the first term of the bound, which comes from adding up the ancestors of the $4$ corners of each of the regions in $L$. Instead of adding up their ancestors up to the root, let us count those ancestors in a finer-grained mode. Consider the $4$ corners of a square of size $\ell_t \times \ell_t$ containing the points in ${\cal P}_t$. Their $\Oh{\log \ell_{p(t)}}$ ancestors of depth over $\lg u - \lg(u/\ell_{p(t)})$ are charged to the node $t$. The higher ancestors, however, cover a square of size at least $2^{\ell_{p(t)}} \times 2^{\ell_{p(t)}}$ by Fact~\ref{fact:basics}, and therefore are also ancestors of some of the $4$ corners of the area of the parent of $t$, $p(t)$. We then do not need to account for those higher ancestors of the corners of the area of $t$. The result follows.
\end{proof}

For example, consider a hierarchical clustering where each cluster lying on a square region of side $\ell$ distributes its points evenly into $c$ sub-clusters lying on squares of side $\ell/s$, for $\log_s u$ levels of clustering. By Theorem~\ref{thm:hier}, the quadtree has $\Oh{n\log s}$ nodes, and its compressed representation uses $\Oh{n\log s}$ bits, which can be $o(n\log u)$.

Due to Fact~\ref{fact:size}, these result also bound the cost of a query on the quadtree of a set of clustered points, because we traverse precisely the quadtree nodes that lead to the output points.

All those results easily generalize to $d$ dimensions, by enclosing each cluster in a hypercube with $2^d$ corners.

\begin{corollary} 
Let ${\cal P}$ be a set of points on the discrete grid $[1,u]^d$ and 
$\mathcal{T}$ be a tree defined as in Theorem~\ref{thm:hier}, except that
now the points ${\cal P}_t$ of every node $t \in \mathcal{T}$ lie on a 
hypercube of side $\ell_t$; those hypercubes need not be disjoint.
Then the quadtree of $\mathcal{P}$ has 
$\Oh{\sum_{t \in \mathcal{T} \setminus \{r\}} 2^d \log \ell_{p(t)} + \sum_{t \in L} n_t \log \ell_t}$ nodes.
\end{corollary}

We can combine these results with the improvements that favor dense clusters \cite{Nav16}, though the formulas are messier: $n_t \log \ell_t$ becomes $n_t\log\min(\ell_t,u/n_i^{1/d})$.

\section{A Compressed Quadtree Representation based on Heavy Paths}
\label{sec:structure}


We describe a new compressed quadtree representation for two-dimensional points which, like the $k^2$-tree, uses $\Oh{1}$ bits per node and supports the basic navigation towards parent and children in $\Oh{1}$ time. In the next section we show that this representation can support queries faster than the $k^2$-tree and, in general, than the standard quadtree representations. We will also generalize our representation to higher dimensions.

\subsection{Data structure}

To store a quadtree, we first replace each internal node by a binary tree of height 2 and remove any node that has no descendant storing a 1. Let $T$ be the resulting binary tree. The number of nodes in $T$ is $1/2$ (when every quadtree node has only one child with with a 1) to $3/2$ (when all quadtree nodes children have 1s) of those in the quadtree. In addition to simplifying our construction, this modification makes quadtrees more practical in higher dimensions~\cite{BDNR14}, which we will also consider at the end of this section.

We then perform a heavy-path decomposition~\cite{ST83} of $T$, as follows.

\begin{definition}
A heavy-path decomposition of $T$ is a recursive decomposition of $T$ into paths called {\em heavy paths}. The first heavy path goes from the root to a leaf, so that if the path contains a node $v$ then it also contains the child of $v$ with the most leaf descendants (breaking ties arbitrarily). Once the first heavy path is defined, its nodes are cut off the tree $T$, leaving a forest of former subtrees of $T$. We then recursively decompose every remaining subtree into heavy paths.
\end{definition}

A well-known property of this decomposition is that every root-to-leaf path in $T$ consists of $\Oh{\log n}$ initial segments of heavy paths. 
In the sequel we call heavy paths simply paths.

\begin{figure}
\begin{center}
\includegraphics[width=0.9\textwidth]{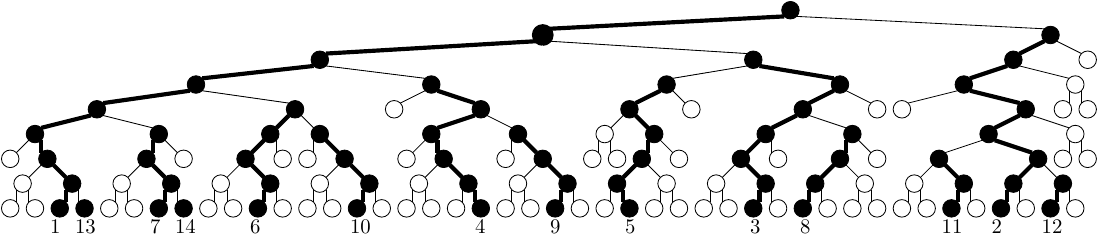}
\caption{The heavy-path decomposition of the binary tree for the example from Figure~\ref{fig:tree}.  Nodes storing 1s are black; nodes storing 0s are shown hollow, and discarded; thick edges belong to heavy paths.  The numbers below the black leaves indicate our path ordering.}
\label{fig:decomposition}
\end{center}
\end{figure}

\begin{example}
Figure~\ref{fig:decomposition} shows the heavy-path decomposition of the binary tree for our example of Figure~\ref{fig:tree}. 
\end{example}

We encode each  path $h$ as a binary string whose 0s and 1s indicate which of $h$'s nodes are left children and which are right children, respectively (considering the root as a left child), in increasing order of their depths. Note that all the  paths end at the same depth, and thus their length plus the depth of their topmost node is the same for all.  

We then sort the set of all those path encodings in decreasing order of their length. Ties between two paths $h$ and $h'$ of the same length are broken as follows: if the topmost nodes of $h$ and $h'$ are $v$ and $v'$, respectively, the paths are ordered in the same way of the paths containing the parents of $v$ and $v'$. Notice that $v$ and $v'$ cannot have the same parent, since they have the same height and the tree is binary. The numbers below the leaves in Figure~\ref{fig:decomposition} indicate how we order the paths in our example.  

Our first structure is a bitvector $H$ that concatenates the encodings of all the  paths, once sorted as described. This bitvector has exactly $|T|$ bits, one representing each node of $T$. We say that the bit \(H [i]\) corresponds to the node $v$ if \(H [i]\) indicates whether $v$ is a left child or a right child.

For each depth \(d < 2\lg u\) (considering the root to have depth 0 and leaves to have depth \(2\lg u\)), we store a bitvector $L_d$ with 1s indicating which nodes at that depth in $T$ have two children. These bitvectors have as many bits as there are internal nodes in $T$.  Figure~\ref{fig:HandL} shows them for our running example.

\begin{figure}[t]
{\small
\begin{eqnarray*}
H   & & \!\!\!\mbox{\tt 000000110\,10010100\,1100010\,110111\,001001\,10010\,1010\,1000\,1110\,1110\,010\,10\,1\,1}\,,\\
L_0 & & \!\!\!\mbox{\tt 1--------}\,,\\
L_1 & & \!\!\!\mbox{\tt -1-------\,0-------}\,,\\
L_2 & & \!\!\!\mbox{\tt --1------\,-0------\,1------}\,,\\
L_3 & & \!\!\!\mbox{\tt ---1-----\,--0-----\,-0-----\,0-----\,0-----}\,,\\
L_4 & & \!\!\!\mbox{\tt ----1----\,---0----\,--1----\,-1----\,-0----\,1----}\,,\\
L_5 & & \!\!\!\mbox{\tt -----0---\,----1---\,---0---\,--0---\,--0---\,-0---\,0---\,0---\,0---\,0---}\,,\\
L_6 & & \!\!\!\mbox{\tt ------0--\,-----1--\,----0--\,---0--\,---0--\,--0--\,-0--\,-0--\,-0--\,-0--\,0--}\,,\\
L_7 & & \!\!\!\mbox{\tt -------1-\,------0-\,-----0-\,----0-\,----0-\,---0-\,--1-\,--0-\,--0-\,--0-\,-0-\,0-}\,; \\
\\
& & P[1..9] = \langle 63, 61, 58, 42, 37, 25, 18, 10, 1 \rangle \\
& & N[1..9] = \langle 12, 11, 10, 6, 5, 3, 2, 1, 0 \rangle
\end{eqnarray*}

\vspace*{-3mm}
}
\caption{The bitvectors $H$ and $L_d$ and the arrays $P$ and $N$ for the tree of Figure~\ref{fig:decomposition}. Dashes and spaces are shown only to indicate how the bits in $L_d$ and $H$ correspond.}
\label{fig:HandL}
\end{figure}

Our final structures are much smaller: an array $P$ of $2\lg u+1$ entries stores in $P[\ell]$ the position in $H$ where the first  path of length $\ell$ (measured in number of nodes) is encoded (or null if there are no  paths of that length). Similarly, an array $N$ stores in $N[\ell]$ the number of paths longer than $\ell$. We give support to perform predecessor queries on $P$ and $N$: $pred(P,i)$ gives the minimum length $\ell$ for which $P[\ell] \le i$. Because we sorted the  paths by decreasing length in $H$, $\ell$ is the length of the  path $H[i]$ belongs to. Similarly, $pred(N,k)$ tells the length $\ell$ of the $k$th path in $H$.

\begin{definition}
Our compressed quadtree representation for $n$ points on a $u \times u$ grid 
has the following components, whose precise contents are defined above:
\begin{itemize}
\item A bitvector $H$ of $|T|$ bits concatenating all heavy paths.
\item Bitvectors $L_d$, for $0 \le d < 2\lg u$, with less than $|T|$ bits in total.
\item Arrays $P$ and $N$, of $\Oh{\log u}$ integers overall.
\end{itemize}
\end{definition}

\begin{theorem}
Our compressed quadtree representation on a set of points in $[1,u]^2$ uses $\Oh{1}$ bits per quadtree node, plus $\Oh{\log^2 u}$ bits.
\end{theorem}
\begin{proof}
Bitvector $H$ and all the $L_d$s take $\Oh{1}$ bits per node in $T$ and, therefore, $\Oh{1}$ bits per node of the original quadtree. The arrays $P$ and $N$, with predecessor data structures, require just $\Oh{\log^2 u}$ further bits.
\end{proof}

\subsection{Navigation}
\label{sec:nav}

In this section we show how the basic parent/child navigation can be supported on our data structures.

\subsubsection{Moving to the parent}
Suppose \(H [i]\) corresponds to node $v$ in $T$. As explained, we obtain the length $\ell$ of the path containing $v$ with $\ell = pred(P,i)$. Further, the path of $v$ is the $k$th (in our ordering) of length $\ell$ and $v$ is the $j$th top-down node in its path, where $k=\lceil (i-P[\ell]+1)/\ell\rceil$ and $j=(i-P[\ell]+1)-(k-1)\ell$. Because the top node in the path of $v$ is at depth $2\log u - \ell$, the depth of $v$ is $d = 2\log u - \ell + j$.

If $j>1$, $v$ is not the topmost node in its path, and then its parent $u$ corresponds to $H[i-1]$. If $j=1$, instead, $u$ belongs to another path. Since $u$ is at depth $d-1$, it is mentioned in $L_{d-1}$. Further, the nodes at depth $d-1$ with a child starting a path are exactly those that have two children (the other child continues the path of its parent). Finally, because we order the paths according to the order of their parent node, it turns out that, since $v$ starts the $k$th path at depth $d$, its parent $u$ is the $k$th node at depth $d-1$ having two children. The position of $u$ in $L_{d-1}$ is then found with $p = select_1(L_{d-1},k)$. 

Because the paths are deployed on $H$ by increasing starting depth (or decreasing length), all the paths having a node at depth $d-1$ precede those that do not. Further, since the nodes in $L_{d-1}$ appear in the same order of $H$, we have that the node at $L_{d-1}[p]$ is the node at depth $d-1$ in the $p$th path of $H$. The length of that path is found with $\ell' = pred(N,p)$, and it is the $k'$ path of length $\ell'$, for $k' = p-N[\ell']$. The position $i'$ of $u$ in $H$ is then computed as $i' = P[\ell']+(k'-1)\ell'+ (d-1) - (2\lg u-\ell')-1$: the paths of length $\ell'$ start at $P[\ell']$, then we have the preceding $k'-1$ paths of length $\ell'$, and in our path we want the node with absolute depth $d'-1$, which we convert to an offset (i.e., relative depth) by subtracting the depth of the first node in the path, $2\lg u - \ell'$.


\begin{example}
Let $v$ be the top node in the ninth path in our ordering (see the first node in the path labeled $9$ in Figure \ref{fig:decomposition}). It corresponds to the underlined position $H[i=50]$. Its parent $u$ is the second node in the fourth path, at $H[i'=26]$. To find $u$ from $i=50$, we first compute $\ell=pred(P,50) = 4$, the length of the path $v$ belongs to. We also compute $k=\lceil (50-42+1)/4\rceil=3$ and $j=(50-42+1)-2\cdot 4 = 1$, so $v$ is the first node in its path, which is the 3rd of length $4$. The depth of $v$ is $d=8-4+1=5$. Since $j=1$, $u$ lies in another path. It is of depth $d-1=4$, and thus mentioned at position $p=select_1(L_4,3)=4$. To find its position in $H$, we compute $\ell'=pred(N,4)=6$, the length of its path, as well as $k'=4-N[6]=1$, so that the path of $u$ is the first of length $6$. We then compute its position $i' = P[6]+0+4-(8-6)-1=26$.
\end{example}

\subsubsection{Moving to a child}
We compute $\ell$, $k$, $j$, and $d$ for $v$ as before.
If the depth $d$ of $v$ is $2\log u$, then $v$ is a leaf; otherwise it has left and/or right children. Further, one of the children of $v$ is at $H[i+1]$: the left child if $H[i+1]=0$ and the right if $H[i+1]=1$.

To determine if $v$ has another child, and where, we must locate $v$ in $L_d$. We compute $r = N[\ell]+k$, the rank of $v$'s path in $H$, thus $v$ is the $r$th node of depth $d$. Therefore, $v$ has another child iff $L_d[r]=1$.

This child is the top node of another path, which starts at depth $d+1$ and thus is of length $\ell''=2\log u - d + 1$. Since there are $s = rank_1(L_d,r)$ nodes of depth $d$ from where new paths start, the child of $v$ is at $H[i'']$, where
$i''=P[\ell'']+(s-1)\ell''$.


\begin{example}
Reversing our previous example, we have for $u$ the values $i=26$, $\ell=6$, $k=1$, $j=2$, and $d=4$.  Since $u$'s depth is $d=4 < 8$, $u$ is not a leaf. One of its children is at $H[26+1]$; since $H[27]=0$, this is the left child of $u$. To find if there is a right one, we compute $r=N[6]+1=4$, so the path of $u$ is the 4th in $H$ and $u$ is the 4th node of depth $4$. Since $L_4[4]=1$, $u$ has a right child. This child starts a path at depth $d+1=5$, of length $\ell''=8-5+1=4$, and it is the 3rd of those because $s=rank_1(L_4,4)=3$. We then find the right child at $H[i'']$, where $i'' = P[4]+2\cdot 4 = 50$, where indeed we find $v$.
\end{example}


\begin{theorem}
Our compressed quadtree representation on a set of 2-dimensional points can move to the parent of a node or to any desired child in $\Oh{1}$ time.
\end{theorem}
\begin{proof}
Moving to the parent or to a child in the quadtree requires a constant number of steps on $T$, each of which is dominated by the time of the predecessor operations on $P$ and $N$. Since these arrays have a logarithmic number of elements, predecessors can be computed in constant time \cite{FW94}. 
\end{proof}

More practically, we note that, on a downward traversal from the root, we always know the length $\ell$ of the current path, and therefore we can perform the operations without the need of predecessor queries. In Section~\ref{sec:membership} we show how heavy paths speed up the specific operations to query quadtrees.

\subsection{Higher dimensions}

In dimension $d$, each quadtree node has $2^d$ children, and the quadtree is still of height $\lg u$. Therefore our binary tree $T$ introduces $d-1$ levels per quadtree edge, reaching height $d\lg u$. Simulating a quadtree move towards the parent or children takes $\Oh{d}$ steps in $T$. Each such step may require predecessor queries on the arrays $P$ and $N$, which now contain $\Oh{d\log u}$ elements. Those predecessor queries can be carried out in time $\Oh{\log\frac{\log d}{\log\log u}}$ \cite{PT06}.

The addition of new nodes in $T$ may increase their number by a factor of $\frac{2^{d+1}-2}{2^d} < 2$ with respect to the number of nodes in the original quadtree (if a quadtree node has $2^d$ children, $T$ adds $2^d-2$ new nodes between the parent and the children). If we consider the number of $1$s in the original quadtree, then $T$ has at most $d$ nodes per $1$ (i.e., a path of length $d$ towards the $1$-child of every quadtree node), and thus $\Oh{d\log u}$ nodes per represented point.

\begin{corollary}
Our compressed quadtree representation on a set of points in $[1,u]^d$ uses $\Oh{1}$ bits per quadtree node (or, alternatively, $\Oh{d\log u}$ bits per point), plus $\Oh{d\log^2 u}$ bits. It can move to the parent of a node or to any desired child in time $\Oh{d\log\frac{\log d}{\log\log u}}$.
\end{corollary}

These space and time factors are similar to what can be obtained on previous compressed quadtree representations \cite{BLN14,Nav16} on high dimensions. Although they can move to parents and children in constant time, their space may grow up to $2^d$ bits per node, that is, exponentially with the dimension, because each node has $2^d$ children and most of them are $0$s. This can be alleviated with a bitvector representation for $B$ that exploits sparsity \cite{OS07}, which recovers the $\Oh{1}$ bits per 1 in the quadtree. In exchange, operation $rank$ takes time $\Oh{d}$, so moving to a child takes time $\Oh{d}$, while moving to the
parent still takes $\Oh{1}$ time.

Although both solutions seem then comparable in terms of space and basic operations, we show next how to leverage the heavy-path representation to support root-to-leaf traversals in time $\Oh{d\log d+\log n}$ instead of $\Oh{d\log u}$. This is particularly relevant for membership queries.

\section{Membership and Range Queries}
\label{sec:membership}

Suppose we want to determine whether the point \((x, y)\) is in the set. The first step is to obtain the Morton code $M$ of the point, by interlacing the bits that describe the integers $x$ and $y$. We can do this in time $\Oh{1/\epsilon}$, for any constant $\epsilon$, with a table using $\Oh{u^\epsilon}$ space. This table does not depend on the data points: for any two chunks of $(\epsilon/2) \lg u$ bits, the table returns their interlacing. We can then find the Morton code of $(x,y)$ by pieces of $\epsilon \lg n$ bits, via $2/\epsilon$ accesses to the table with the consecutive pieces of $(\epsilon/2)\lg u$ bits of $x$ and $y$.

We now enter the binary tree $T$ from the root, using the successive bits of $M$ to
decide whether to go left or right. That is, each bit of $M$ corresponds to an edge to follow.
Instead of processing $M$ bit by bit and descending in $T$ edge by edge, however, we descend path by path. 

We first determine the prefix of the first path (the one starting at the root) that we must follow. For this sake, we compute $m = \LCP(M[d+1..],H[p+1..])$, where $d=0$, $p=1$, and $\LCP(X,Y)$ is the length of the longest common prefix between bitstrings $X$ and $Y$ (we start from $p+1$ because the first bit of the first path, $H[1]$, is spurious, whereas $H[2]$ refers to the first edge). Note that the length of the first path, $H[p..]$, is $\ell = 2\lg u+1$, with $\ell-1$ edges.

If $m=\ell-1$, then $M[d+1..]$ matches the whole path starting at $H[p]$, and then we know that the point is stored in the quadtree. If not, then $M[d+1..]$ shares its first $m$ edges with the path starting at $H[p]$, matching up to node $H[p+m]$, but not $H[p+m+1]$ (e.g., if $m=0$ and $p=1$ then $M$ matches only the root node). We must then determine if $H[p+m]$ has two children and, if so, move to the other child. This is done as described in Section~\ref{sec:nav}.
If $H[p+m]$ has only one child, then $(x,y)$ is not in the set. Otherwise, letting $H[p']$ be the other child of $H[p+m]$, we know that $H[p']$ starts a path of length $\ell'=\ell-m-1$, with $\ell'-1$ edges. We then update $p \gets p'$, $d \leftarrow d+m+1$, $\ell \leftarrow \ell-m-1$.

This process is repeated until we find $(x,y)$ or determine it is not in the set of points. Since we switch to another descendant heavy path at each step in our process, we perform $\Oh{\log n}$ steps \cite{ST83}. Algorithm~\ref{alg:member} gives
the pseudocode.

\begin{algorithm}[t]
\caption{{\bf Membership}($x$, $y$)}
\label{alg:member}

$M[1..2\lg u] \gets$ Morton code of $(x,y)$; \\
$s \gets 1$; \\
$\ell \gets 2\lg u+1$; \\
$p \gets 1$; \\
$d \gets 0$; \\
\While{true}
   { $m \gets \textrm{LCP}(M[d+1..],H[p+1..])$; \\
     \lIf{$m=\ell-1$} {\Return yes}
     $r \gets N[\ell]+s$; \\
     \lIf{$L_{d+m}[r]=0$} {\Return no}
     $s \gets rank_1(L_{d+m},r)$; \\
     $\ell \gets \ell-m-1$; \\
     $p \gets P[\ell]+\ell \cdot (s-1)$; \\
     $d \gets d+m+1$;
   }
\end{algorithm}

Note that we always know the length $\ell$ of the path we are navigating, and therefore moving to a child requires only $\Oh{1}$ time. Just the $rank$ functionality on the bitvectors, without using $select$ nor predecessor queries, is needed. We can also compute $m=\LCP(X,Y)$ in $\Oh{1}$ time if $|X|$ and $|Y|$ are $\Oh{\log u}$ (we apply $\LCP$ on $H[p+1..]$, but it suffices to consider only the first $2\lg u$ bits of that suffix). With $Z = X~\texttt{xor}~Y$, the $m$ highest bits become $0$ and the $(m+1)$th becomes $1$. We then use a constant-time technique to find the highest $1$ in $Z$ \cite{Knu09}. We can also compute $\LCP$ using tables of size $u^\epsilon$, as before.


\begin{example}
To perform a membership query for \((6, 9) = (0110_2,1001_2)\) in our quadtree of Figure~\ref{fig:tree} (the shaded cell), we first interlace the bitstrings to obtain the path label, $M = 10010110$, and then use it to traverse the path-decomposed tree $T$ of Figure~\ref{fig:decomposition}. We first try to match $M[1..]$ with the longest path, of length $9$, from $H[2..]=00000110\ldots$ (see Figure~\ref{fig:HandL}). The common prefix is of length $m=0$, so we cannot go past the root by that path. We then see that the root has another child, which is $H[10]$, starting a path of length $9-0-1=8$ (our algorithm computes $r=1$, $s=1$, $\ell=8$, $p=10$, and $d=1$, continuing because $L_0[1]=1$). Since $m=\LCP(M[2..],H[11..])=5$, we can advance up to $H[15]$, where $M$ wants to go right ($M[7]=1$) but the path goes left ($H[16]=0$). We then find that $H[15]$ has another child, $H[61]$, which starts a path of length $8-5-1=2$ (our algorithm computes $r=2$, $s=1$, $\ell=2$, $p=61$, and $d=7$, continuing because $L_6[2]=1$). We finally compute $m=\LCP(M[8..],H[62])=1$, so we have arrived at a leaf ($m=\ell-1$) and report that the point exists.
\end{example}

In the worst case, we must traverse $\Oh{\log n}$ paths along this process. This contrasts with the $\Oh{\log u}$ time needed with the classical representation, showing that our structure should be faster on sparse points sets. Further, we need fewer path switches in the way to isolated points. The next theorem shows that the membership time indeed improves on those points.

\begin{theorem}
\label{thm:membership}
With the help of a constant table using $u^\epsilon$ space (for any constant 
$\epsilon>0$), our compressed quadtree representation supports a membership 
query for \((x, y)\) in $\Oh{\log n}$ time. Further, the time is \(\Oh{\min_g \{\log (u / g) + \log k_g\}}\), where $k_g$ is the number of points in $\mathcal{P}$ within distance $g$ of \((x, y)\).
\end{theorem}

\begin{proof}
The $\Oh{\log n}$ bound follows from the heavy path decomposition. Further,
any ancestor $v$ of $(x,y)$ of depth at least \(2 \lg (u / g) + 2\) in $T$ covers a subgrid of size at most $g/2 \times g/2$, whose points that are then at distance at most $(g/2)\sqrt{2} < g$ from $(x,y)$. Thus, $v$ covers at most $k_g$ points of $\cal P$.  It follows that the path from $v$ to the deepest ancestor $w \in T$ of $(x,y)$ consists of $\Oh{\log k_g}$ initial segments of heavy paths.  To see why, consider that if we ascend from $w$ to $v$, every time we move from the topmost node in one heavy path to its parent in another heavy path, the number of leaf descendants in the subtree below us at least doubles.  Since the path from the root to $v$ has length $\Oh{\log (u / g)}$, the path from the root to $w$ consists of $\Oh{\log (u / g) + \log k_g}$ initial segments of heavy paths.
\end{proof}

The following corollary, which combines Theorems~\ref{thm:hier} and~\ref{thm:membership}, suggests that our structure should be particularly suited to applications in which points are highly clustered (e.g., towns) but queries are chosen uniformly or according to a different distribution (e.g., seismic activity). 

\begin{corollary} 
\label{cor:hier-query}
Let ${\cal P}$ be a set of points on the discrete grid $[1,u]^2$ and
$\mathcal{T}$ be a tree defined as in Theorem~\ref{thm:hier}.
Let $\mathcal{T}_\ell$ be the set of nodes at level $\ell$ in $\mathcal{T}$. Then a membership query for $(x,y)$ takes $\Oh{\min_\ell \max_{t \in \mathcal{T}_\ell} \log(u/d({\cal P}_t,(x,y)))}$ time, where $d({\cal P}_t,(x,y))$ is the minimum distance between a point in ${\cal P}_t$ and $(x,y)$.
\end{corollary}
\begin{proof}
Let $m_\ell = \max_{t \in \mathcal{T}_\ell}\log(u/d({\cal P}_t,(x,y))) = 
\log(u/\min_{t \in \mathcal{T}_\ell} d({\cal P}_t,(x,y)))$. Let us define
$g = (1/2) \min_{t \in \mathcal{T}_\ell} d({\cal P}_t,(x,y))$, so $k_g=0$
and $m_\ell = \Oh{\log(u/g)}$. By Theorem~\ref{thm:membership}, the cost
of the membership query is then $\Oh{\log(u/g)+\log k_g} = \Oh{m_\ell}$. This 
bound holds for every level $\ell$, so the time is $\Oh{\min_\ell m_\ell}$.
\end{proof}


\subsection{Range queries}

In order to output all the points in $\mathcal{P} \cap R$ given a query region $R = [x_1,x_2] \times [y_1,y_2]$, we first traverse via heavy paths towards the lowest ancestor of $R$ in the quadtree. For this sake, we compute $m = \min(\LCP(x_1,x_2),\LCP(y_2,y_2))$ and define $x$ and $y$ as the first $m$ bits of $x_1$ and $y_1$, respectively. We then interlace $(x,y)$ into a bitstring $M$ of length $2m$, and traverse towards $M$ in the quadtree as described in the main part of this section. The node $v$ we arrive at is the ancestor of all the points in $R$. We now traverse edge by edge towards all the descendants of $v$ using the method described in Section~\ref{sec:nav}, in constant time per edge traversed, avoiding to enter into nodes whose area does not intersect $R$, and reporting all the leaves found.

The node $v$ can be very high in the tree, even for small regions, and thus this method is no faster in the worst case than the classical one. We expect, however, it to be faster in many cases when the query region is small. We inherit the
refined bounds for classic quadtrees \cite[p.~361]{Nav16}.

\begin{corollary}
Our compressed quadtree representation outputs $Q = \mathcal{P}\cap R$, for a rectangle $R$ of size $p \times q$, in time $\Oh{p+q+\log u+|Q|(1+\log(pq/|Q|))}$.
\end{corollary}

\subsection{Higher dimensions}

In dimension $d$, the description of the point sought, $(x_1,\ldots,x_d)$, has $d\lg u$ bits, and it might not fit in a computer word of size $\Oh{\log u}$. We can still use the precomputed table of size $u^\epsilon$ described at the beginning of this section to obtain the bitstring $M$ (of length $d\lg u$) in time $\Oh{(1/\epsilon)d\log d}$, as follows. Build a binary tree where the root represents all the $d$ coordinates, $1,\ldots,d$. Its left child represents the odd positions of the parent, $1,3,5,\ldots$ and the right child the even positions, $2,4,6,\ldots$. This division, taking odd and even positions, continues until the leaves represent only one dimension $1 \le i \le d$ and store the bitstrings $x_i$. Now, bottom up, every internal node merges the bitstrings of its two descendants by chunks of $(\epsilon/2)\lg u$ bits,  until the root obtains $M$. It is easy to see that the tree has $\lg d$ levels and that the total work per level is $\Oh{d/\epsilon}$.

Once we obtain $M$, the membership query proceeds as for two dimensions. The only difference is that a single $\LCP$ query may take time $\Oh{d}$, because the prefix may coincide in $m=\Oh{d\log u}$ bits. However, the sum of all those $m$ values along the search is also $\Oh{d\log u}$, because we advance in $M$ by $m+1$ positions each time. Therefore, the time to descend to the leaf $(x_1,\ldots,x_d)$ or to determine it does not exist is $\Oh{d+\log n}$. Note that we do not require predecessor queries to determine membership.

\begin{corollary}
With the help of a constant table using $u^\epsilon$ space (for any constant $\epsilon>0$), our compressed quadtree representation supports a membership query for \((x_1,\ldots,x_d)\) in \(\Oh{d\log d + \log n}\) time.
\end{corollary}

Our finer results can be similarly extended to $d$ dimensions; we leave them
as exercises to the reader.

\section{Practical Optimizations and Implementation Variants}
\label{sec:practice}

For the creation of the path labels, we use a precomputed table of 256 entries to compute the interleaving byte-wise, together with some arithmetics to build the final path. For the 3-dimensional case we have also used an implementation based on magic numbers. In practice, the computation of the initial path has negligible effect on the total query times.

The query algorithms described in previous sections always use a top-to-bottom traversal of the conceptual tree $T$, which leads to a number of practical optimizations.  A first practical choice, already mentioned, is the adjustment of the traversal algorithms to keep track of the current depth, thereby avoiding the need for predecessor structures in $P$ and $N$. In this section we describe other practical variants that can reduce the space usage of our structure.

A first significant space improvement can be obtained by removing information in $H$ that can be deduced during top-down traversals. Bitvector $H$ stores, for each path, a bitstring representing its nodes, marking whether each is a left or a right child. During top-down traversal, we always start at the beginning of the first path, and whenever we switch to a new path we always start at its beginning. Note that when switching paths, the first bit of the new path can be inferred, since it is the opposite of the next bit in the current path. Indeed, in the membership query of Section~\ref{sec:membership} we always skip that first bit, $H[p]$, and compare $M[d+1..]$ with $H[p+1..]$. Therefore, we can remove the first bit of each path in $H$ and still perform top-down traversals on the tree. Since all the paths are shortened in the same way, the navigational properties remain the same and only minor changes are required. Overall, we save one bit per path, or which is the same, per point in $\cal P$.

Another improvement in space can be obtained by noting that there are only $n$ $1$s across all the bitvectors $L_d$, at the starting node of each path. In contrast, their total length can be up to $2n\lg u$ (i.e., one path in $T$ per point), getting closer to that maximum on sparse datasets. We can then use for the bitvectors $L_d$ a compressed bitvector representation \cite{RRR} that supports constant-time access and $rank$ queries but uses less space when the bitvector has many more $0$s than $1$s. With that representation, the whole set of bitvectors $L_d$ fits within $\Oh{n\log\log u}$ bits. In practice this representation is slower, though.

Finally, we can also apply to our structure the matrix-vocabulary compression applied on the last levels of the quadtree \cite{BLN14}, described at the end of  Section~\ref{sec:comprquad}. We can trim $T$ a few levels above the last one, and use DACs to replace the removed levels by a matrix vocabulary and a DAC-encoded sequence of matrix identifiers. All the query algorithms remain the same, though they stop at a smaller depth $d'$. Once this depth is reached, we find the corresponding submatrix, and perform single-cell access or range queries over the submatrix in the same ways as the classical compressed quadtree \cite{BLN14}.

\section{Experimental Evaluation}
\label{sec:experiments}

\subsection{Experimental framework}

We tested the performance of our solution on real datasets from different domains. We consider grids extracted from geographic information systems (GIS), social networks (SN), Web graphs (WEB) and RDF datasets (RDF). 

\begin{itemize}
\item The datasets \dblp and \enwiki are network data corresponding to the social network datasets dblp-2011 and enwiki-2013, provided by the Laboratory for Web Algorithmics\footnote{{http://law.di.unimi.it}}~\cite{BoVWFI,BRSLLP}.
\item Collections \indo and \uk are obtained from Web graph crawls, from the indochina-2004 and uk-2002 datasets provided by the Laboratory for Web Algorithmics.
\item We build three datasets storing geographic information by processing the Geonames dataset, which stores over 9 million locations, discretizing them on a grid. We build three different datasets (\giss, \gism, \gisd) by varying the resolution of the grid.
\item We build three datasets storing RDF-based grids, by parsing the DBPedia dataset\footnote{http://wiki.dbpedia.org/Downloads351}. RDF stores triples (S,P,O) that represent labeled edges in a graph, where the predicate P represents the label. We partition the dataset by predicate, so each individual element can be regarded as a binary grid, and select three different datasets, \rdfs, \rdfm, and \rdfd, with significantly different number of points in the grid.
\end{itemize}

These datasets aim at testing the performance of our technique on a wide variety of real-world applications. The selection of GIS-based and RDF-based datasets also aims at providing insights on its relative performance depending on the sparsity of the data, which is a key element for our structure. Table~\ref{tab:datasets} describes the main characteristics of the studied datasets, including the grid size (all grids are square, of size $u \times u$) and the number $n$ of points in the grid.

\begin{table}
\centering

\begin{tabular}{ l | l | r r }
 File  & Type & Grid size ($u$) & Points ($n$) \\
 \hline
\dblp       & SN  &     986,324 &   6,707,236 \\
\enwiki     & SN  &   4,206,785 & 101,355,853 \\
\hline
\indo  & WEB &   7,414,866 & 194,109,311 \\
\uk         & WEB &  18,520,486 & 298,113,762 \\
\hline
\giss   & GIS &  67,108,864 &   9,335,371 \\
\gism      & GIS &   4,194,304 &   9,328,003 \\
\gisd    & GIS &     524,288 &   9,188,290 \\
\hline
\rdfs  & RDF &  66,973,084 &     138,303 \\
\rdfm   & RDF &  66,973,084 &   7,936,138 \\
\rdfd   & RDF &  66,973,084 &  98,714,022 \\
\hline
\end{tabular}
\caption{Description of the datasets used in our experiments}
\label{tab:datasets}
\end{table}

We compare our representation with the \kt \cite{BLN14}, the best known compressed quadtree representation, which has been shown to achieve very good compression on most of those domains, especially on Web graphs and RDF data. We use two different implementations of the \kt: \ktplain is a direct implementation of the structure, with all the bitmaps stored in plain form, and using $k=2$ in all levels of the tree; \ktdac is an enhanced version that applies a number of improvements over the basic approach: it uses $k=4$ in the first 6 levels of decomposition and $k=2$ in the remaining levels; also, DACs are used to replace the last 3 levels of the tree by submatrices of size $8 \times 8$.

We also compare our representation with path-decomposed tries (\pdt) \cite{GO14}. We use two of the configurations proposed by the authors that provide a reasonable space-time tradeoff: the centroid hollow monotone-hash technique (\pdth) and the centroid compressed trie, where labels are compressed using RePair (\pdtrp). Note that \pdt is designed to represent string dictionaries, and only supports membership queries. In order to transform the point grids into collections of strings suitable for \pdt, we use the Morton code for each individual point in the collection and build the \pdt representation of the collection of Morton codes. Membership queries are directly translated into \pdt operations, but queries involving rows/columns or ranges are not specifically supported and are therefore transformed into a number of membership queries.

We test four different implementations of our proposal, \hpqt, considering two main variables. First, bitvectors $L_d$ can be stored in plain form (\hpqtp) or compressed with the so-called RRR technique \cite{RRR} (\hpqtR). Second, we may use our basic implementation, with all  paths stored completely, or use the DAC-based compression of the submatrices in the lower levels. This compression leads to two further variants, \hpqtpdac and \hpqtRdac.

We implemented the \hpqt variants in C++, using LibCDS 2 \footnote{https://github.com/fclaude/libcds2} to provide the bitvector implementations used. Both \kt implementations are provided by the authors and implemented in C. PDT is implemented in C++, and obtained from the original author's repository\footnote{https://github.com/ot/path\_decomposed\_tries}. All implementations were compiled using GCC with full optimization enabled. Experiments were executed on a machine with Intel Xeon E5-2470@2.3GHz (8 cores) CPU, and 64GB of RAM. The operating system was Debian 9.8 (kernel 4.9.0-8-amd64).

\subsection{Space usage}

\begin{table}
\footnotesize
\centering
\begin{tabular}{ l | r r r | r r r  | r r}
 & & & & \multicolumn{3}{|c|}{$^{\DAC}$} & \multicolumn{2}{c}{\pdt} \\
Dataset & \hpqtp & \hpqtR & \ktplain & \hpqtp & \hpqtR & \kt & hollow & RP \\
\hline
\dblp	& 11.62	& 9.23	& 10.76	& 10.08	& \textbf{8.88}	& 9.84 & \underline{9.00} & 25.43\\
\enwiki	& 17.56	& 13.49	& 16.96	& 15.01	& \underline{13.37}	& 14.66 & \textbf{9.11} & 31.07\\
\hline
\indo	& 3.29	& 2.92	& 2.57	& 1.28	& \underline{1.27}	& \textbf{1.22} & 7.51 & 16.10 \\
\uk	    & 4.04	& 3.73	& 3.30	& 2.08	& \textbf{2.02}	& \underline{2.04} & 7.99 & 17.04\\
\hline
\giss	& 44.19	& 29.66	& 44.01	& 38.32	& \underline{28.48}	& 38.02 & \textbf{8.23} & 51.15\\
\gism	& 30.61	& 21.28	& 30.10	& 25.42	& \underline{20.72}	& 24.83 & \textbf{8.22} & 40.64\\
\gisd	& 17.37	& \underline{13.05}	& 16.55	& 13.85	& 13.21	& 13.17 &\textbf{ 8.14} & 29.81 \\
\hline
\rdfs	& 45.01	& 30.35	& 45.69	& 39.81	& \underline{29.63}	& 46.98 & \textbf{11.06} & 57.36 \\
\rdfm	& 11.19	& 9.02	& 9.80	& 7.36	& \underline{7.09}	& \textbf{6.93} & 9.28 & 24.30 \\
\rdfd	& 31.94	& 22.22	& 31.61	& 27.28	& \underline{21.54}	& 26.93 & \textbf{9.31} & 43.42 \\
\hline
\end{tabular}
\caption{Space required by all implementations (in bits per point). We put in bold the best and underline the second best space for each dataset.}
\label{tab:space}
\end{table}

In this section we compare the compression performance of \hpqt with \kt variants. Table~\ref{tab:space} displays the compression obtained, in bits per point, in all the test datasets. Let us first focus on the comparison between \hpqt variants and the \kt. The results show that \hpqtR variants achieve better compression than the \kt in almost all the datasets. The \kt only obtains the best compression results in \indo and \rdfm, two datasets where the differences between representations are not very high in general. Plain versions, \hpqtp, are still slightly larger than the equivalent \kt, both with basic representations and with DAC. 

Let us compare the compression obtained by \pdt variants, displayed in the last two columns of Table~\ref{tab:space}. The space usage of \pdt follows patterns completely different from the other alternatives: \pdth is very consistent, using 8--10 bits per point in all the datasets, whereas \pdtrp requires much more space, ranging from 16 to 57 bits per point depending on the dataset. The consistency of \pdth makes it much more efficient than \hpqt to represent datasets with no clear regularities in the points, such as clustering. For instance, in the \giss dataset, \pdth is 3--5 times smaller than the \hpqt variants, and in \rdfs it is roughly 3--4 times smaller. However, in the Web graph datasets, \pdth is far from the compression offered by \hpqt or \kt variants, becoming 2--4 times larger than our proposal. Note that the space partitioning of \hpqt and \kt is expected to work well on sparse grids with clustered points, whereas \pdt does not explicitly consider any of this.

Overall, the results show that \hpqt outperforms \kt in space in almost all cases, achieving good compression especially on Web graphs. Alternatives like \pdth, which do not exploit point regularities, outperform both \hpqt and \kt in space on datasets where the points are distributed more randomly, for example on GIS. We recall, however, that \pdt is not designed for representing point grids, as it does not support range queries. The next sections complement this analysis by testing the query performance of all these representations.

\subsection{Membership queries}

We now test the performance of our technique for membership queries, which check whether a given point exists in the grid or not. We test separately for empty and filled cells, and perform a third test on isolated filled cells. For each dataset and query type, we build a collection of 100,000 query points. For empty and filled cells we select these points at random, whereas for isolated cells we select the 100,000 points that are farthest away from their closest neighbor. We run each full query set 100 times in each dataset and measure the average.

\begin{figure}[t]
\begin{center}
\includegraphics[angle=-90,width=0.48\textwidth]{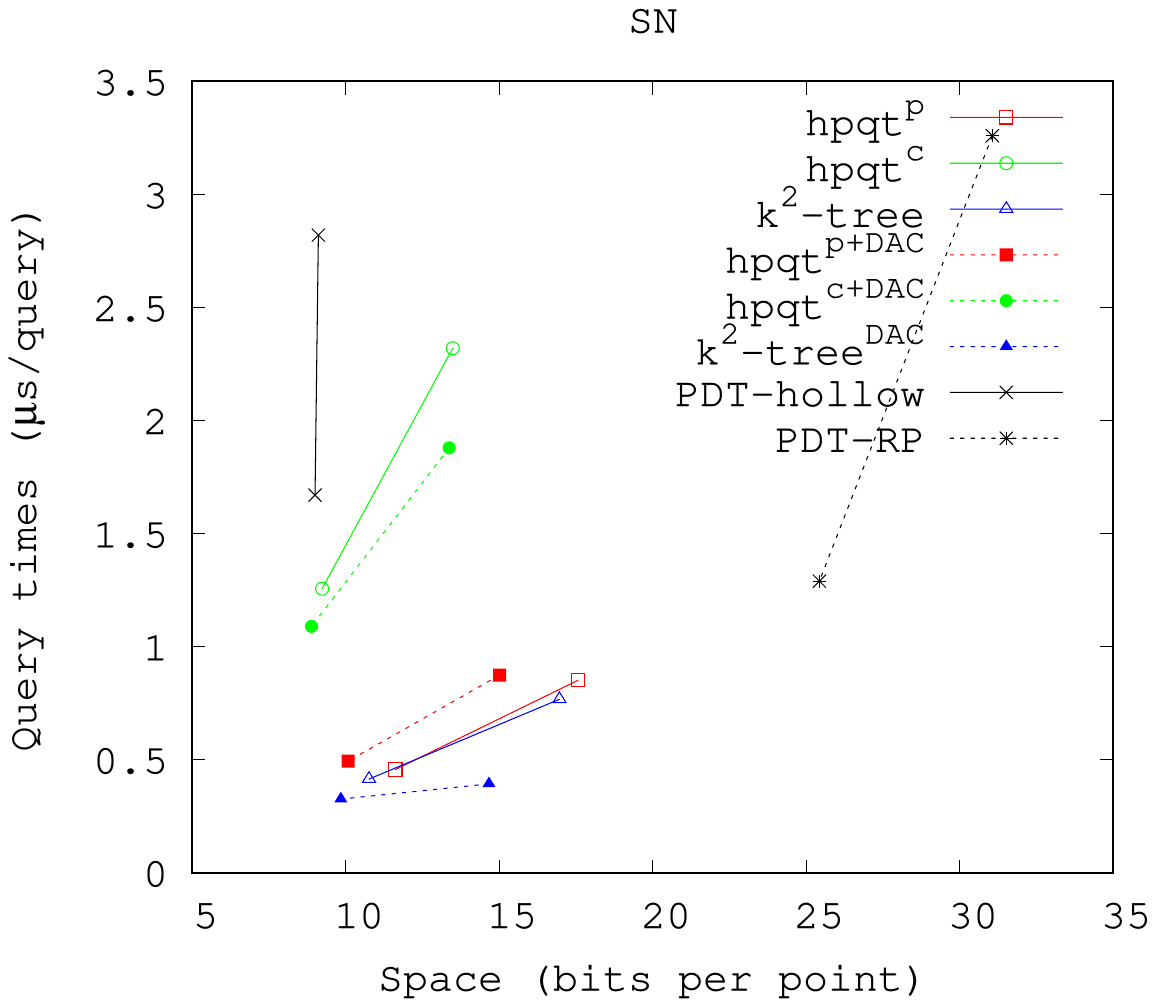}
\includegraphics[angle=-90,width=0.48\textwidth]{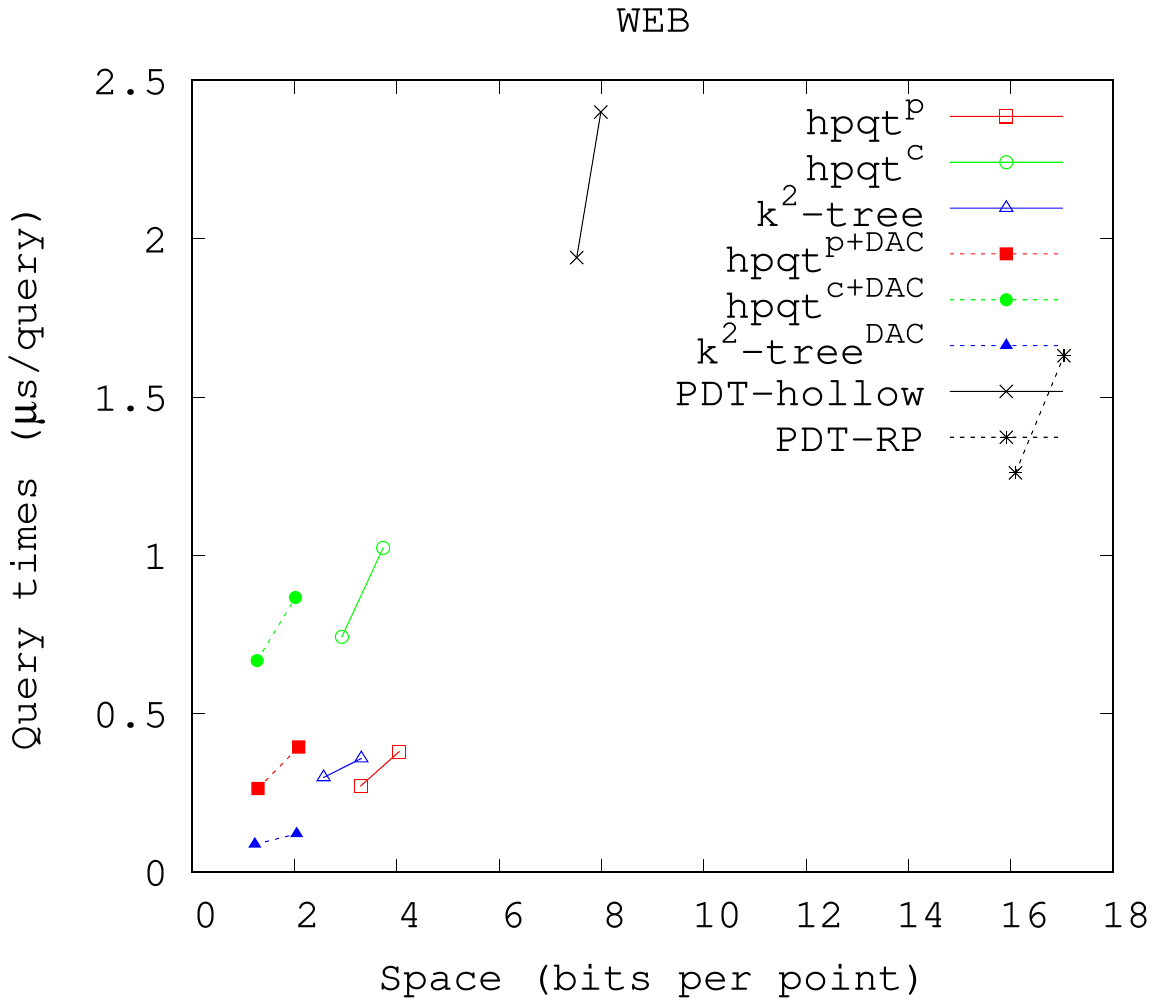}
\includegraphics[angle=-90,width=0.48\textwidth]{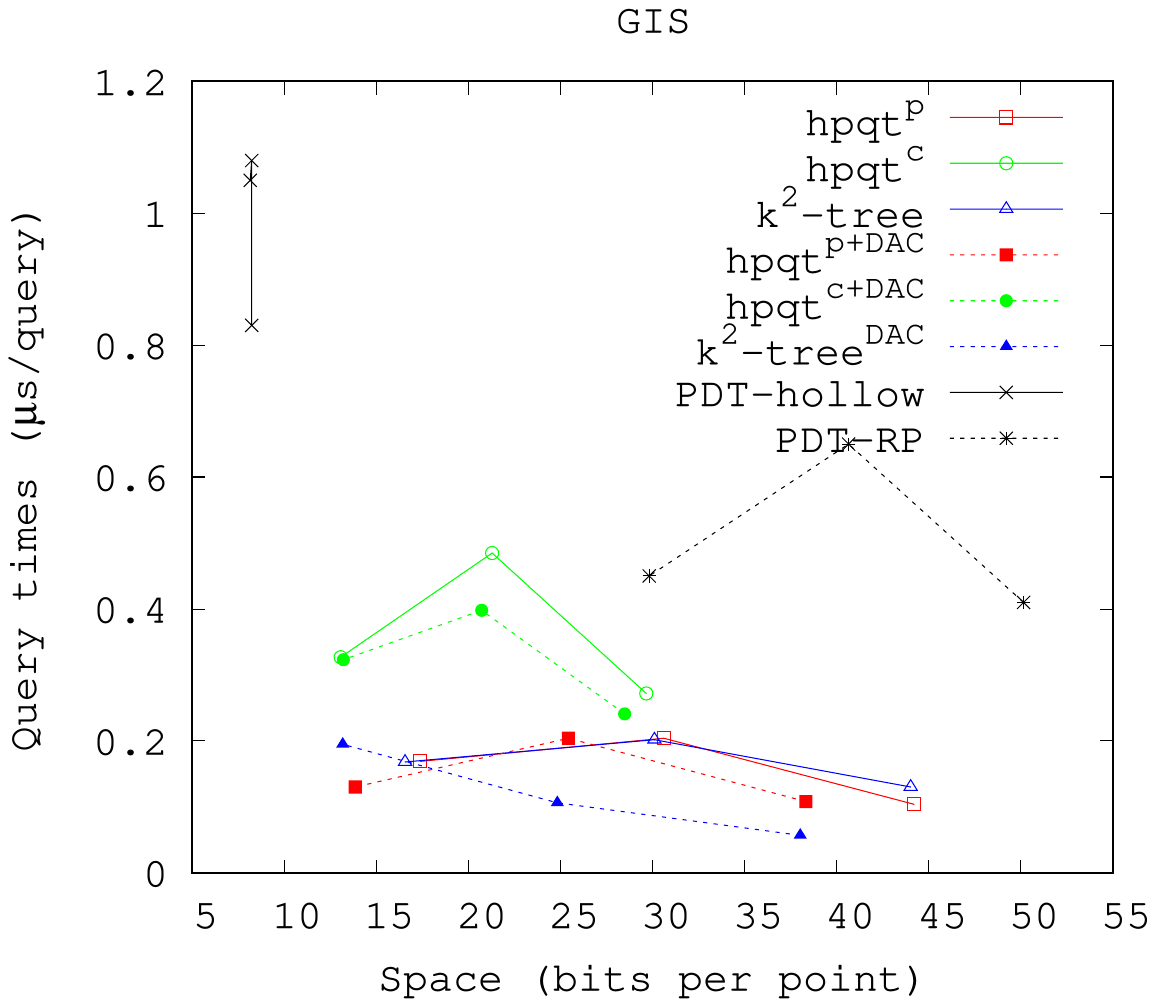}
\includegraphics[angle=-90,width=0.48\textwidth]{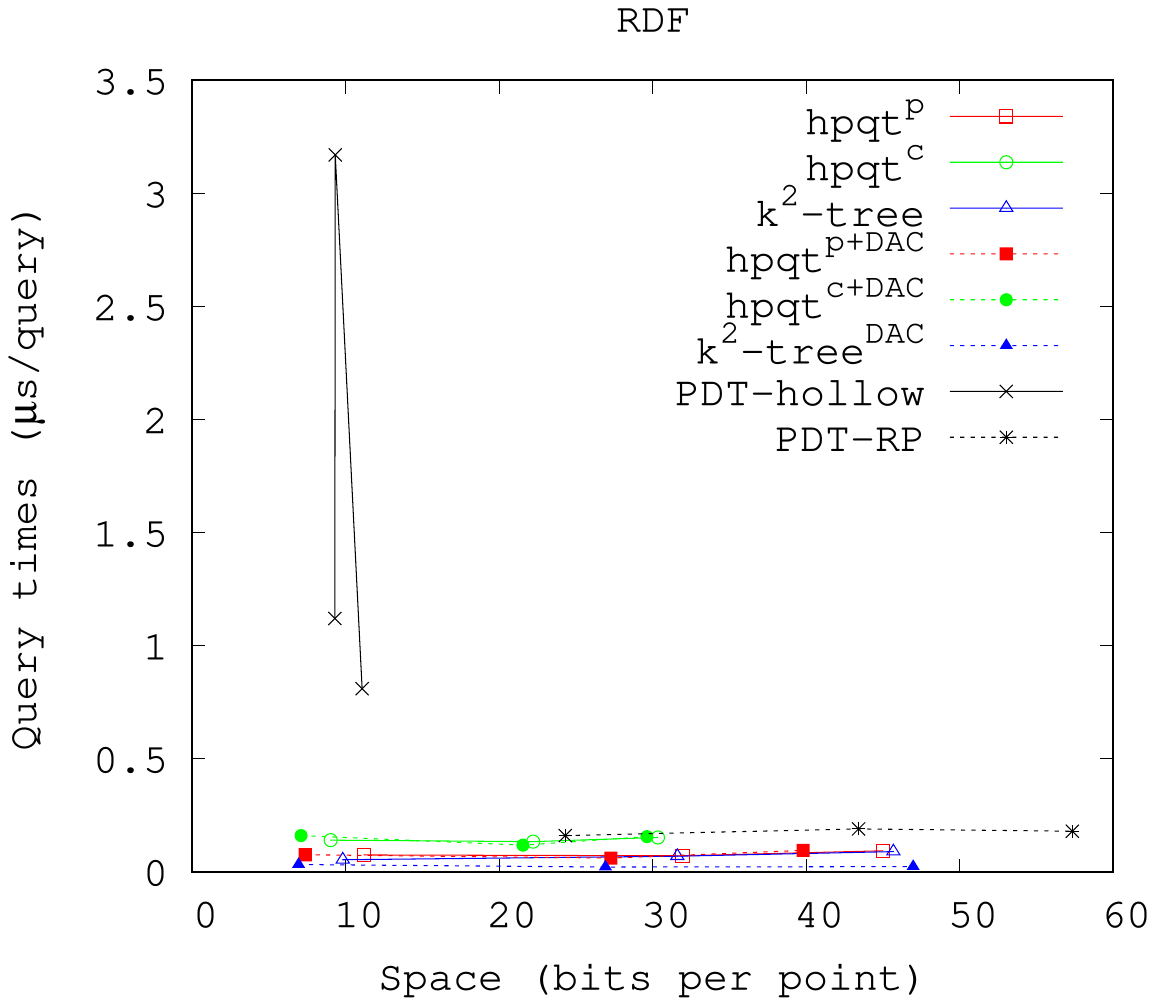}
\end{center}
\caption{Query times of membership queries for empty cells. Times are in $\mu$s/query. Datasets are grouped by family, and lines join the points on different datasets of the same family: SN includes, from left to right, \dblp--\enwiki; WEB includes \indo--\uk; GIS includes \gisd--\gism--\giss; RDF includes \rdfm--\rdfd--\rdfs.}
\label{fig:timesempty}
\end{figure}

Figure~\ref{fig:timesempty} displays the result of membership queries for empty cells on all the datasets. The datasets are grouped by family, and we describe the tendency of each method in the datasets of the same family using lines. One first general conclusion is that the $\DAC$ variants are smaller and faster than those representing all the nodes in bitvectors. In general, the best variants by far are always \hpqtpdac and \ktdac, the latter being always slightly smaller and almost always faster, by a smaller or a larger margin, with the exception of the dense GIS datasets. All \kt and \hpqt variants improve in general on sparser or more clustered point sets,  because isolated empty cells tend to be higher in the quadtree, but the \kt exploits this effect better because its search cost is directly proportional to the depth of the leaf sought. Our compressed variants, \hpqtR, are significantly slower than the plain ones, \hpqtp, and provide relevant space-time tradeoffs only on the sparser GIS and RDF datasets, where points distribute uniformly and then the $L_d$ bitvectors tend to have about $2\lg u$ bits per $1$. In many cases \pdth offers very attractive space, though it is also significantly slower. The variant \pdtrp is never competitive.

%

\begin{figure}[t]
\begin{center}
\includegraphics[angle=-90,width=0.48\textwidth]{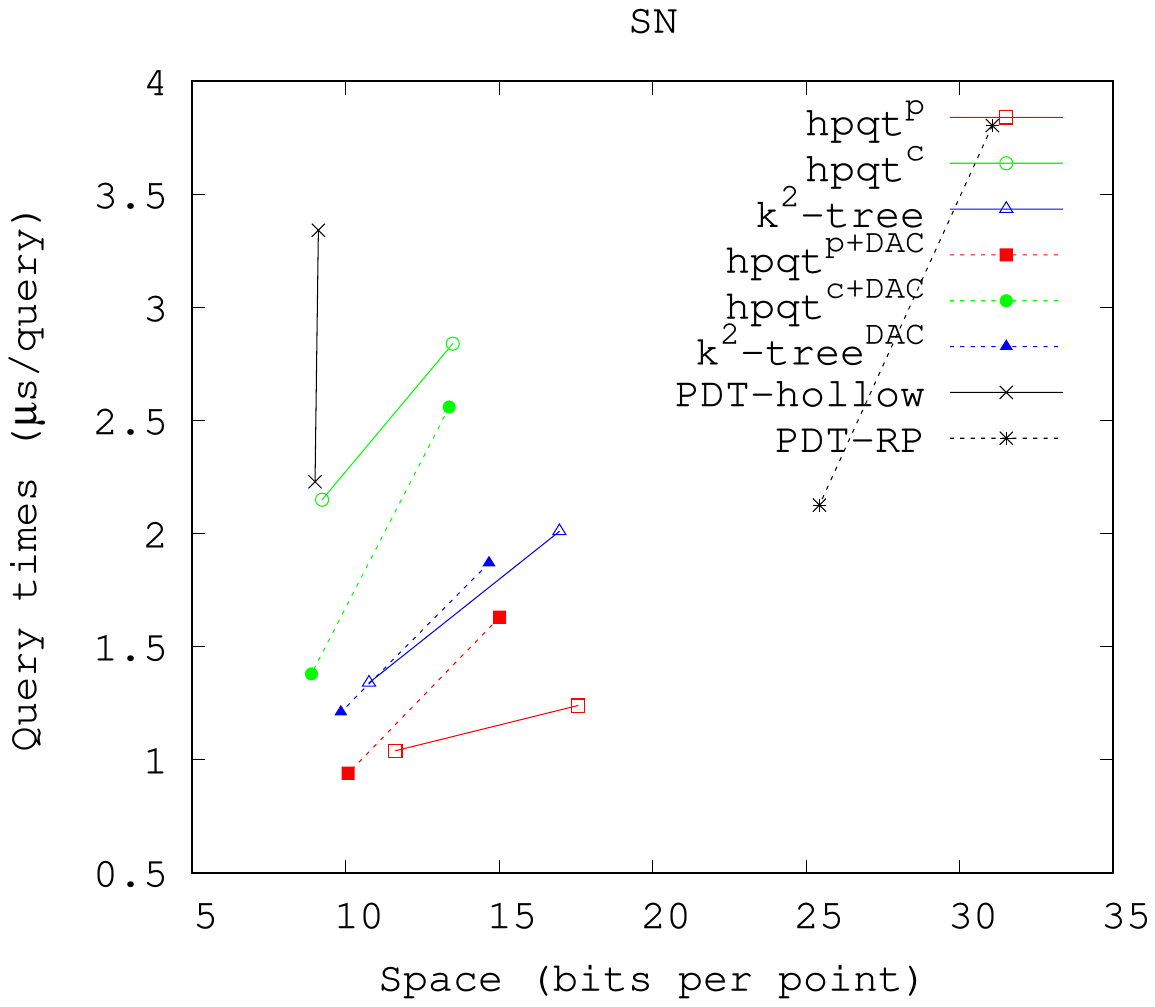}
\includegraphics[angle=-90,width=0.48\textwidth]{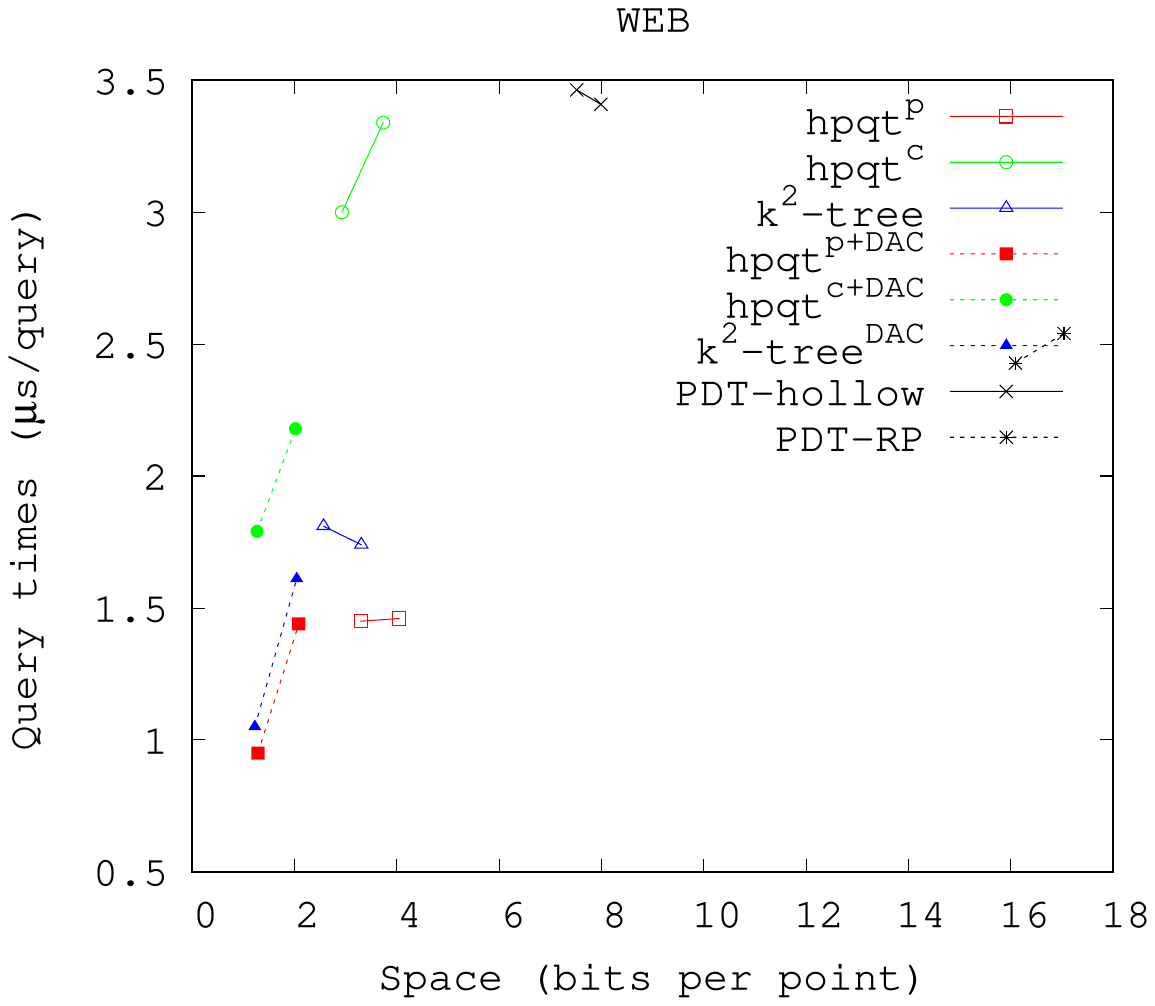}
\includegraphics[angle=-90,width=0.48\textwidth]{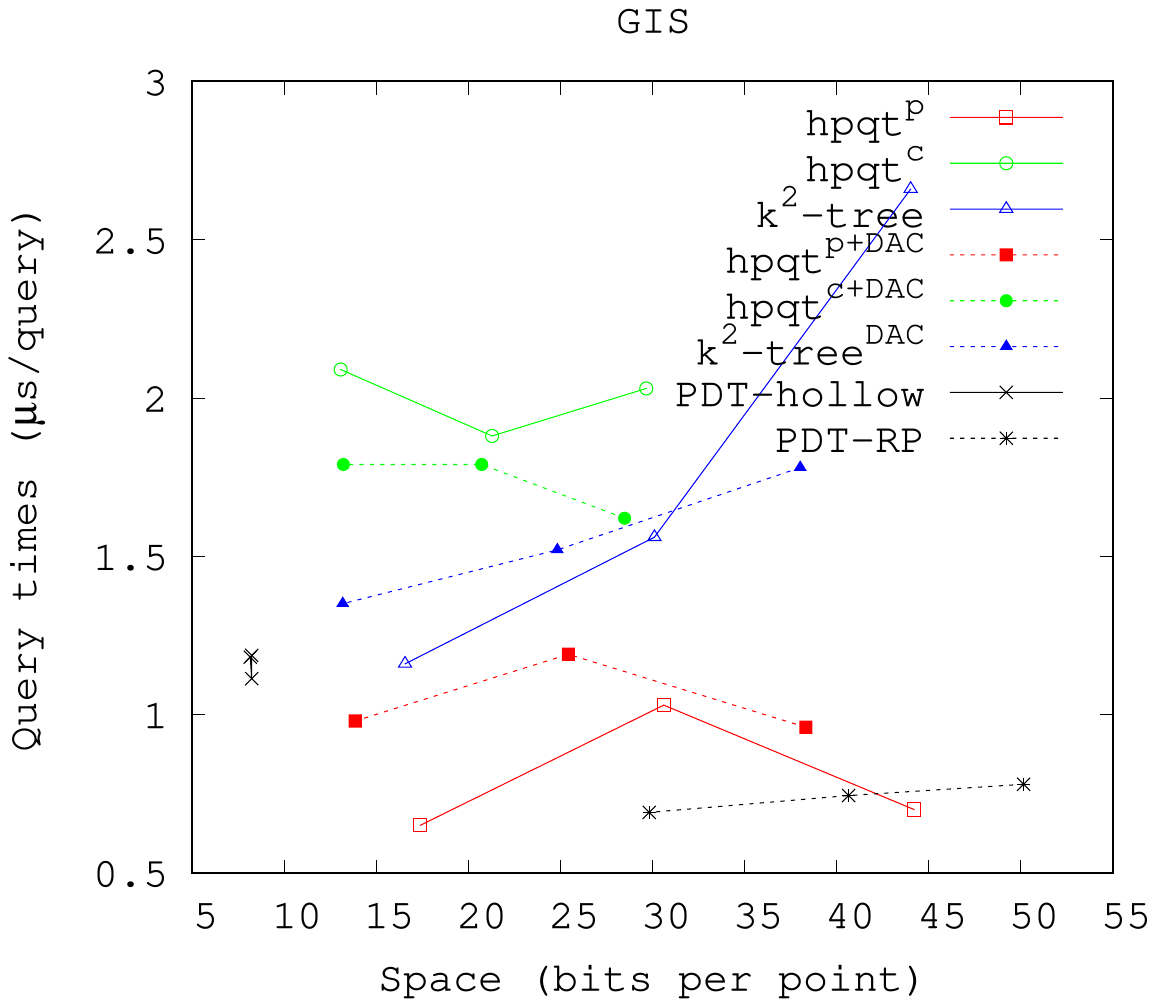}
\includegraphics[angle=-90,width=0.48\textwidth]{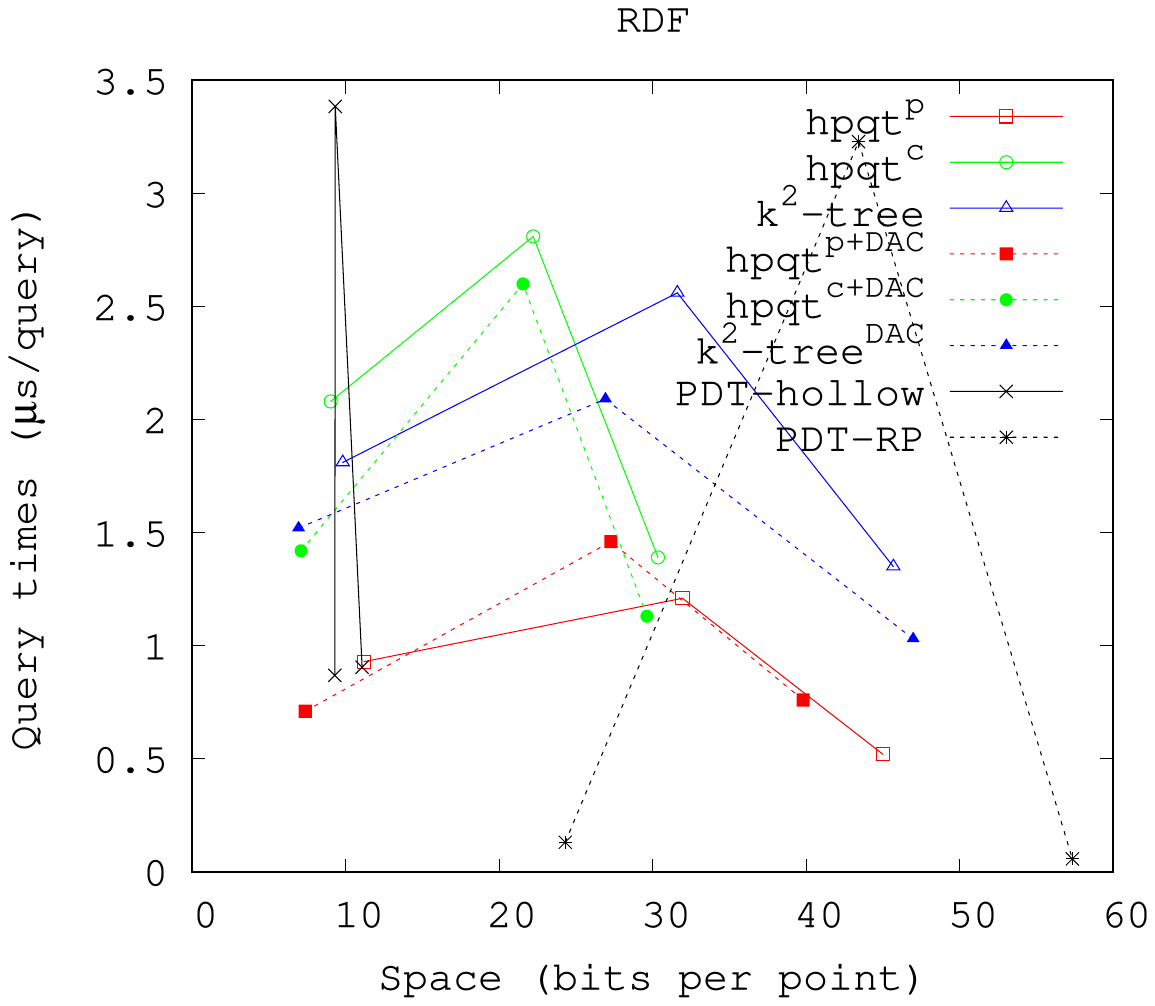}
\end{center}
\caption{Query times of membership queries for filled cells. Times are in $\mu$s/query.}
\label{fig:timesfilled}
\end{figure}

Figure~\ref{fig:timesfilled} displays the result for filled cells, following the same grouping of datasets used in Figure~\ref{fig:timesempty}. For these queries, \hpqtp and \hpqtpdac become clearly faster than the \kt variants, while using similar space. In the sparser GIS and RDF datasets, even the slow compressed variants, \hpqtR and \hpqtRdac, outperform the \kt both in space and time. On the other hand, \pdth and \pdtrp are also much more competitive, especially on the GIS and RDF datasets, where \pdth is still the smallest by far (except on \rdfm) but now its speed is much more competitive. In turn, \pdtrp is by far the fastest in many cases, though it is considerably larger in general. Note that the query times of \pdt do not change significantly with respect to Figure~\ref{fig:timesempty}, whereas accessing filled cells is much more expensive for \hpqt and especially for \kt, because filled leaves are always in the deepest level. 

\begin{figure}[t]
\begin{center}
\includegraphics[angle=-90,width=0.48\textwidth]{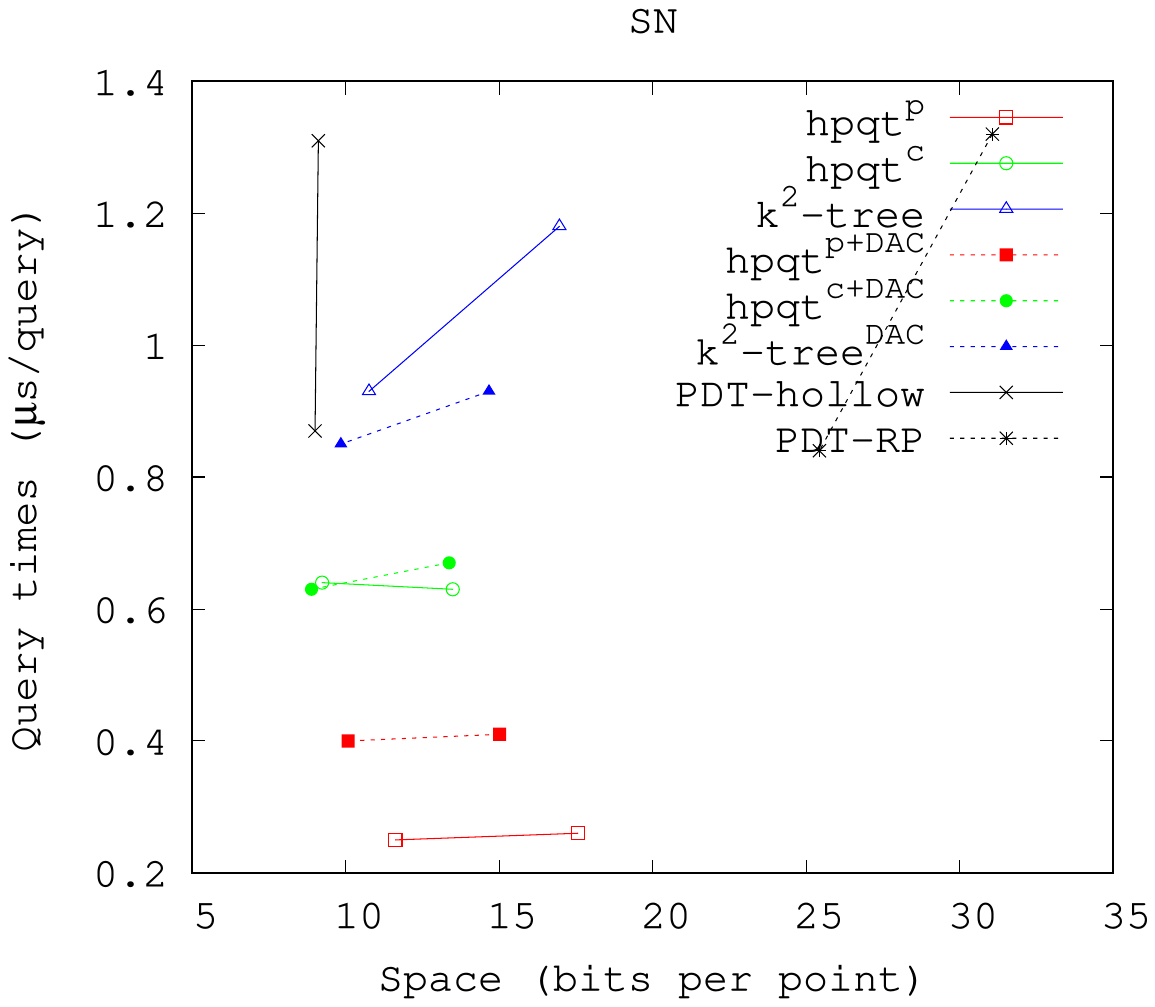}
\includegraphics[angle=-90,width=0.48\textwidth]{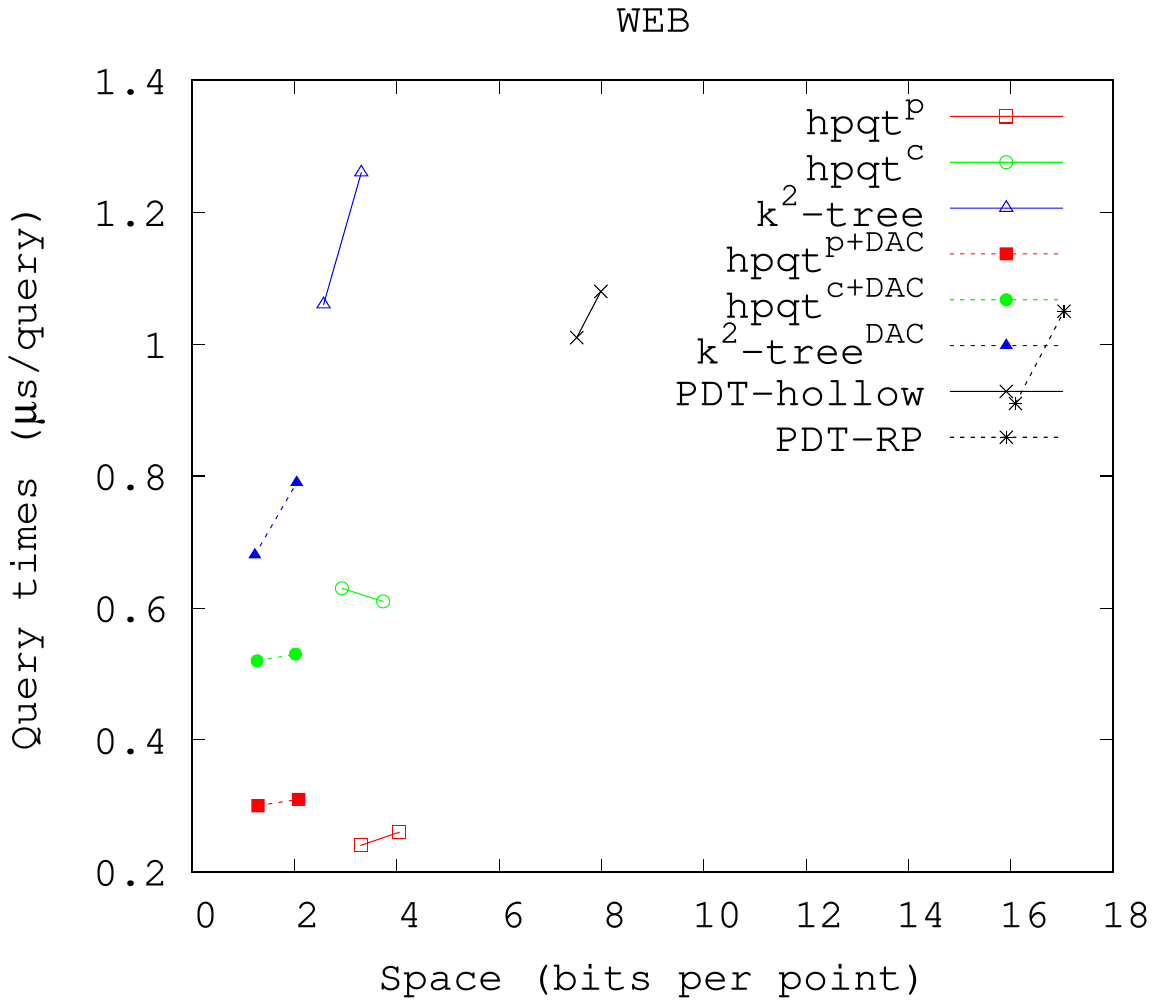}
\includegraphics[angle=-90,width=0.48\textwidth]{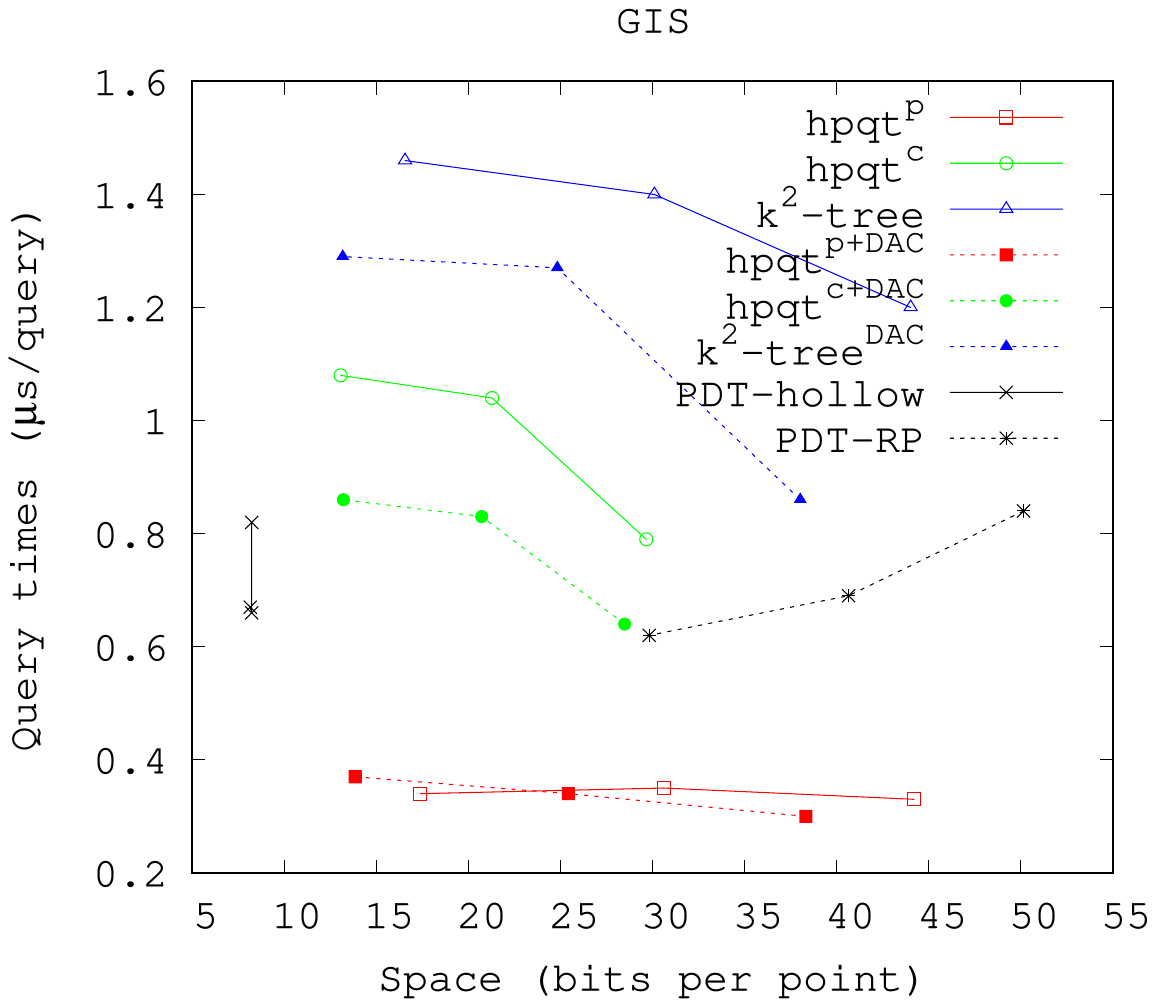}
\includegraphics[angle=-90,width=0.48\textwidth]{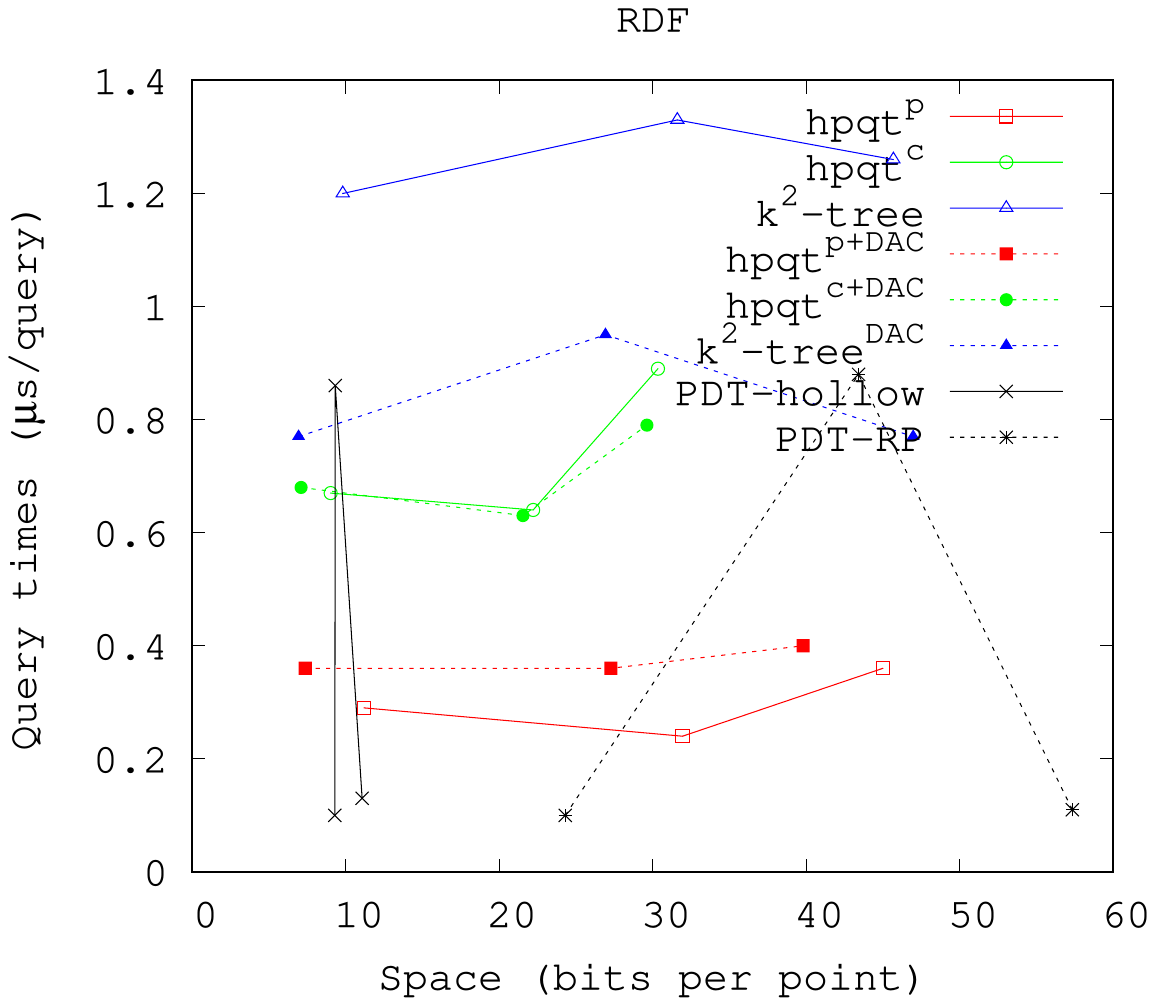}
\end{center}
\caption{Query times of membership queries for isolated filled cells. Times are in $\mu$s/query.}
\label{fig:timesisolated}
\end{figure}

Finally, Figure~\ref{fig:timesisolated} displays the query times for isolated filled cells, where our structure excels, becoming 2--3 times faster than for random filled cells. Now even the bitvector-compressed \hpqt variants are faster than the \kt. 
As in the previous cases, \pdth provides a competitive space-time tradeoff in some datasets, especially on the sparser RDFs, and \pdtrp is sometimes the fastest (though also the largest), but the margin is much narrower than before. 

\subsection{Range queries}

We now consider range queries, which ask for all the points in a defined window of the grid. 
For these experiments, we selected fixed square window sizes (4, 16, 64, 256, 1024), and for each window size and dataset we built sets of 1,000 random window queries.


\begin{figure}[t]
 \centering
     \includegraphics[angle=-90,width=0.49\textwidth]{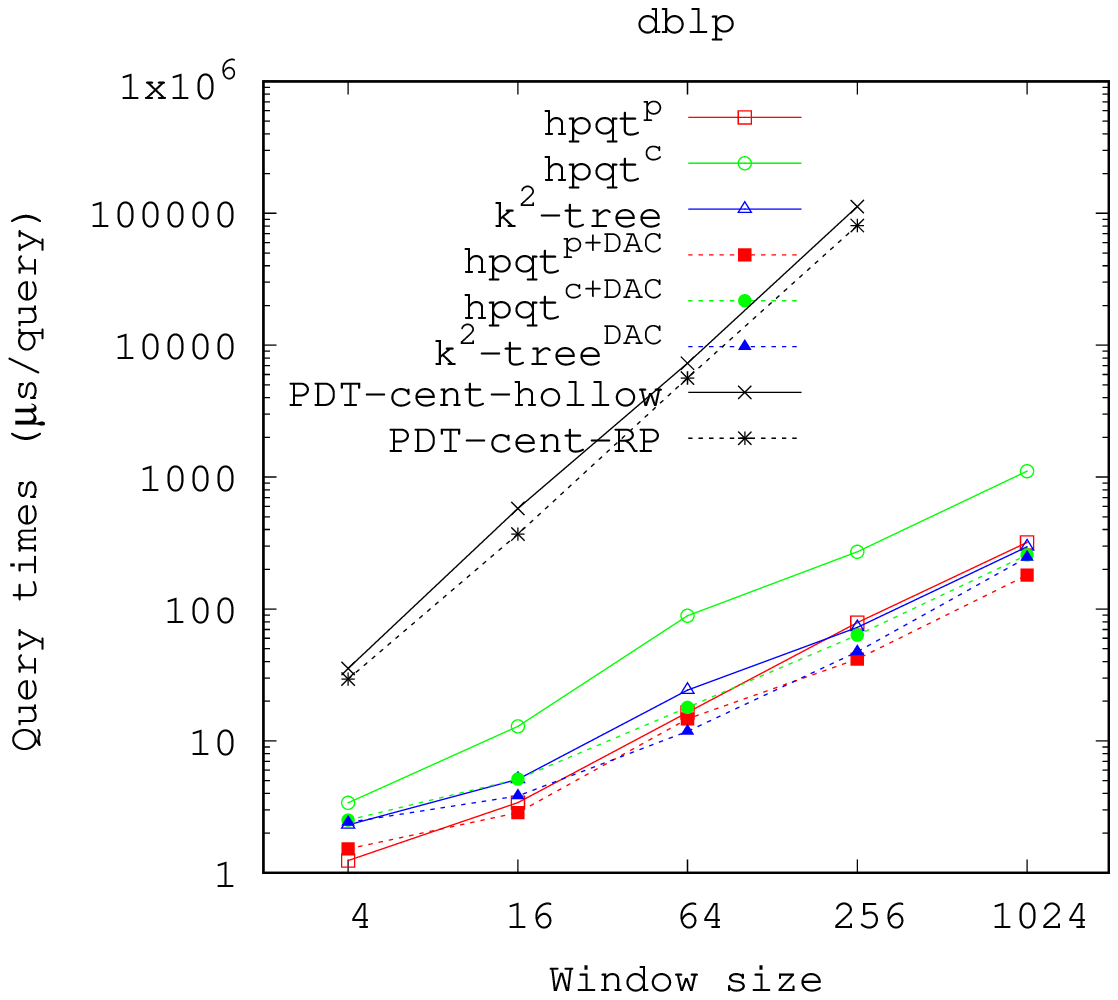}
    \includegraphics[angle=-90,width=0.49\textwidth]{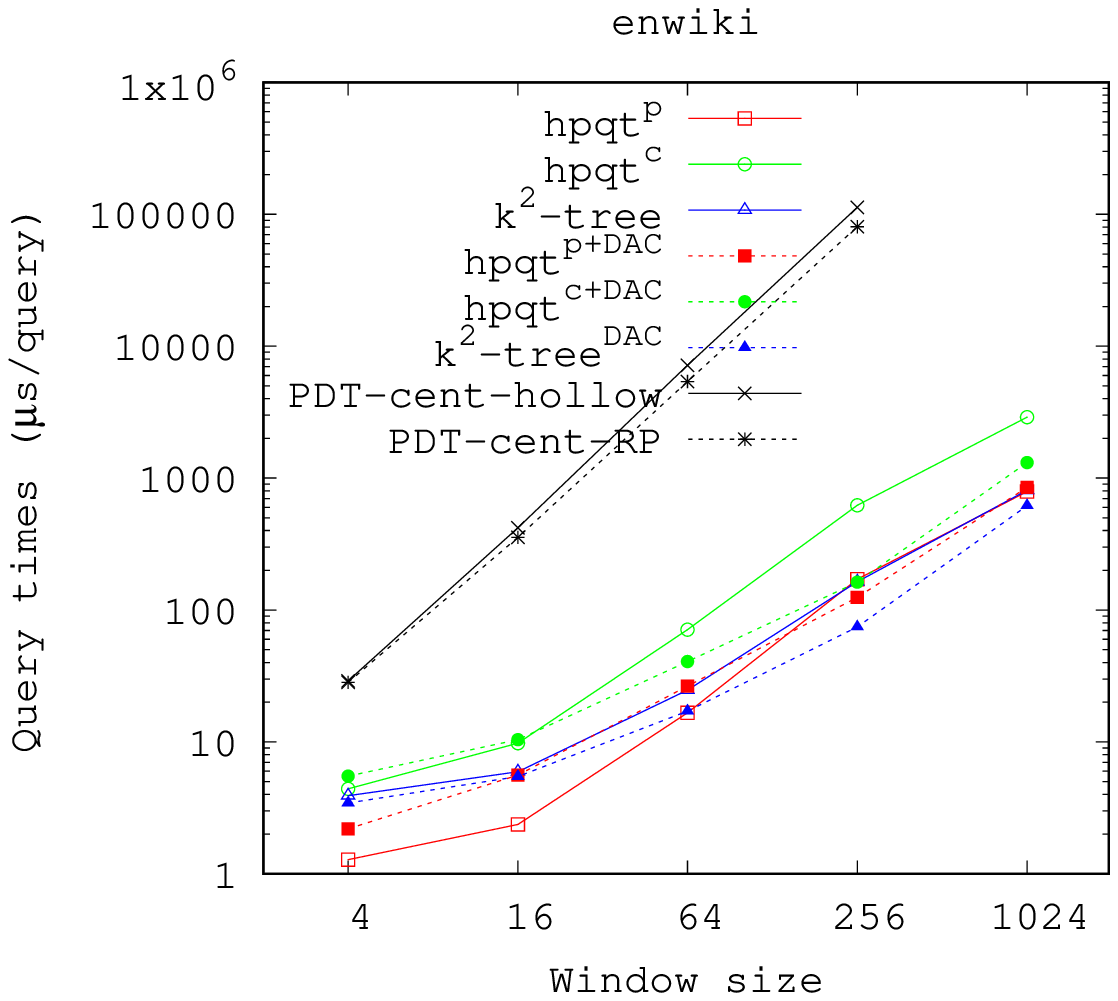}
    \includegraphics[angle=-90,width=0.49\textwidth]{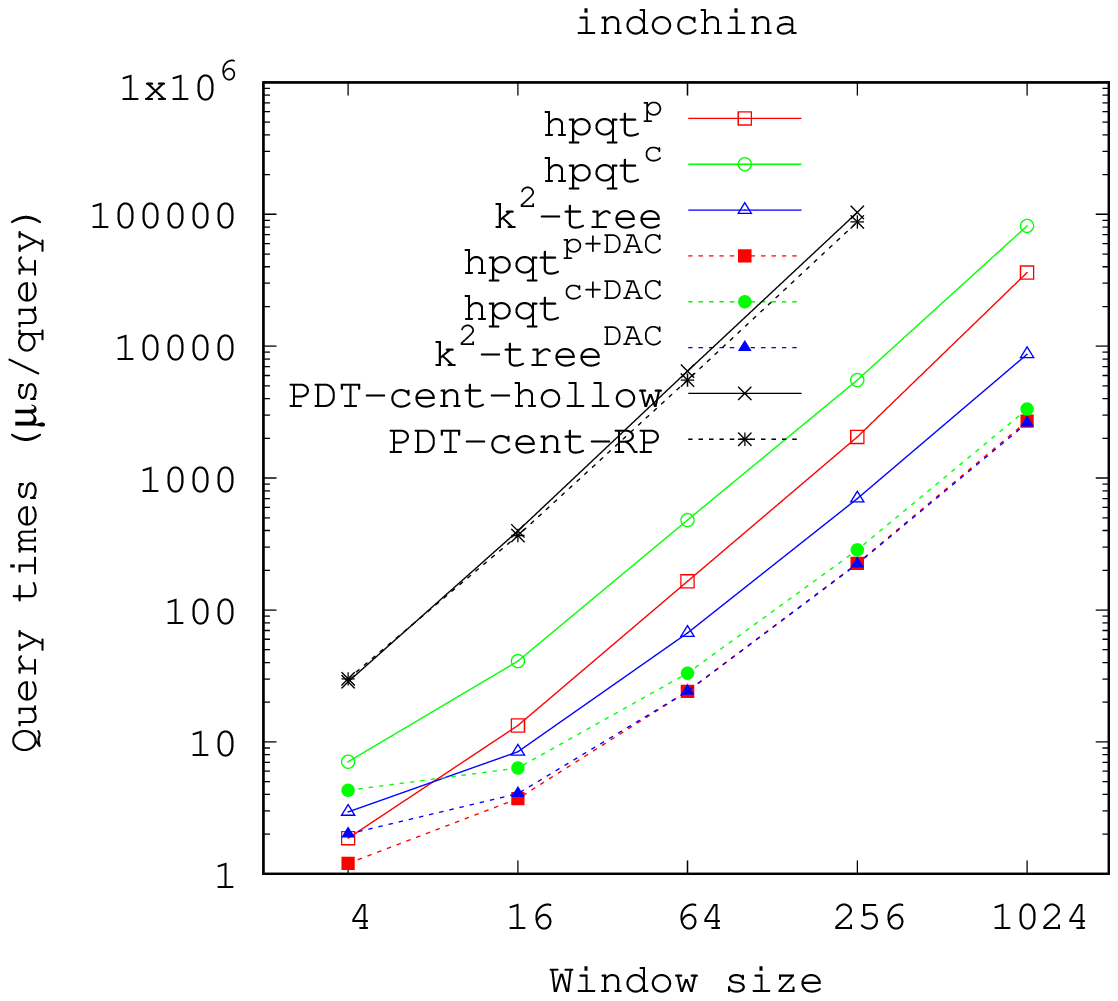}
    \includegraphics[angle=-90,width=0.49\textwidth]{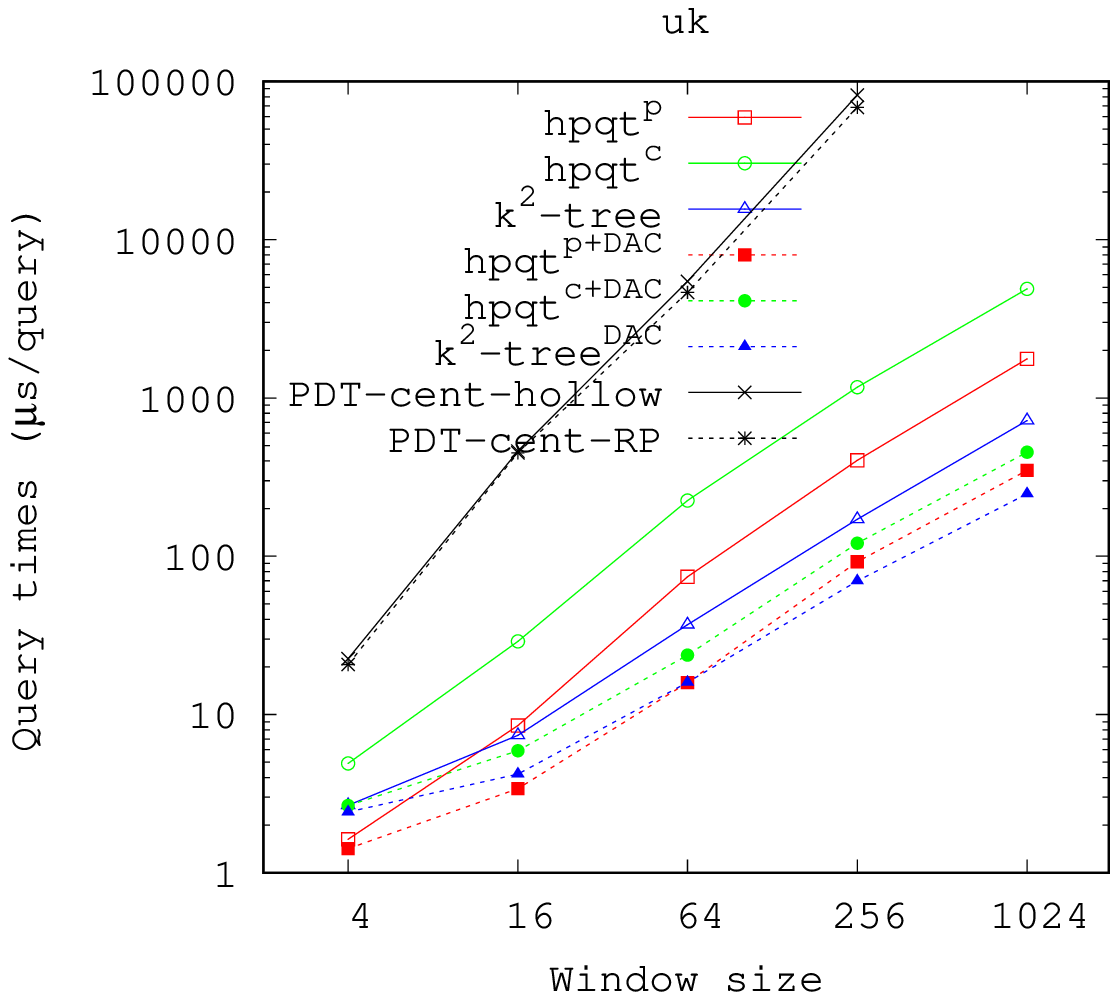}
    \caption{Query times for window queries, with varying window size, for SN (top) and WEB (bottom) datasets. Results are in $\mu$s/query. Log-scale is used in both axis.}
  \label{fig:rangeSW}
\end{figure}

Figure~\ref{fig:rangeSW} displays the query times obtained on the social networks and Web graphs, with varying window size. In this type of queries, \hpqtp, \hpqtpdac, and \ktdac are always the fastest. On small windows, \hpqt is more efficient to reach the deepest node that contains the window, whereas on larger windows \ktdac takes over, generally by a small margin. Note that the bitvector-compressed variant \hpqtRdac is always competitive in time as well. Finally, note that \pdth and \pdtrp are orders of magnitude slower even on the smallest windows, because they do not support range queries and we must resort to individual searches of all the possible points in the query window. 


\begin{figure}[t!]
 \centering
    \includegraphics[angle=-90,width=0.49\textwidth]{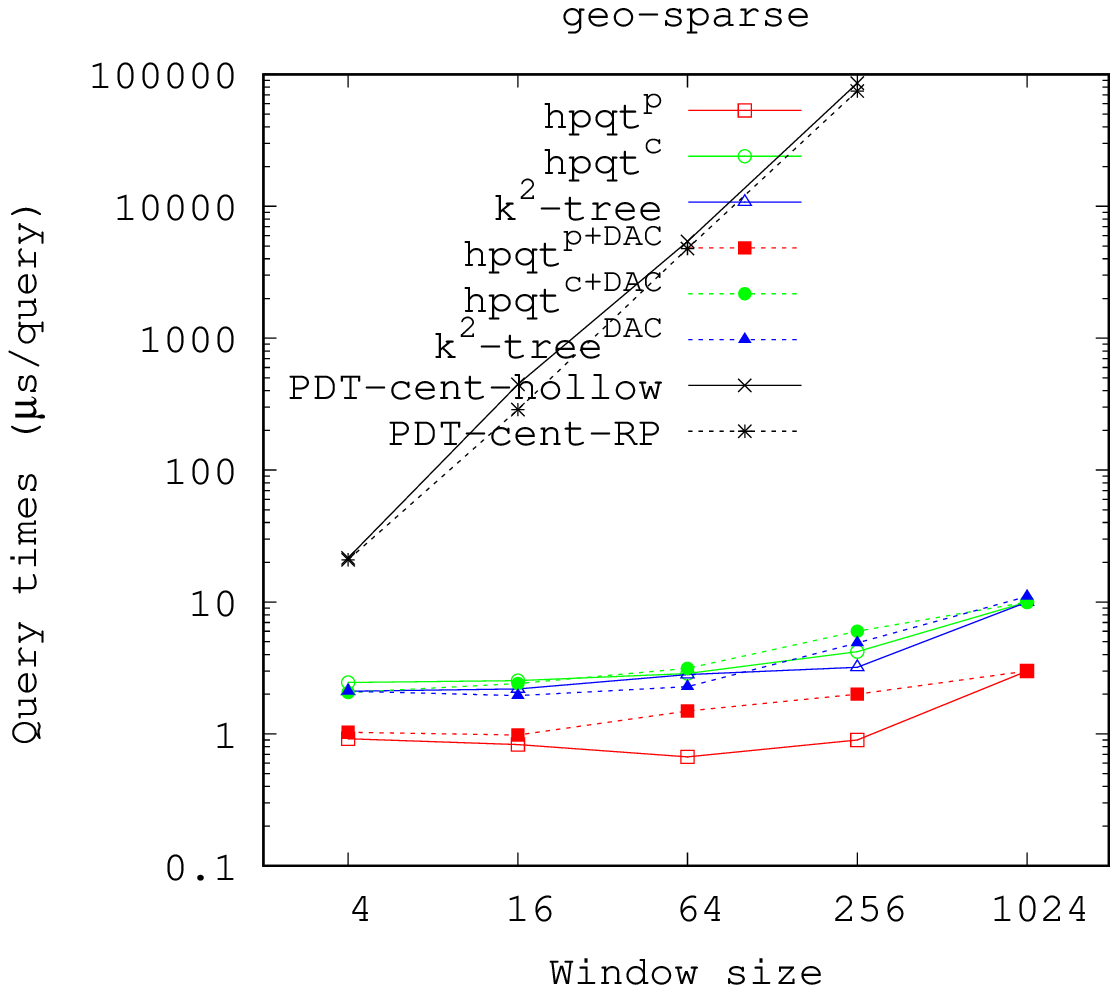}
    \includegraphics[angle=-90,width=0.49\textwidth]{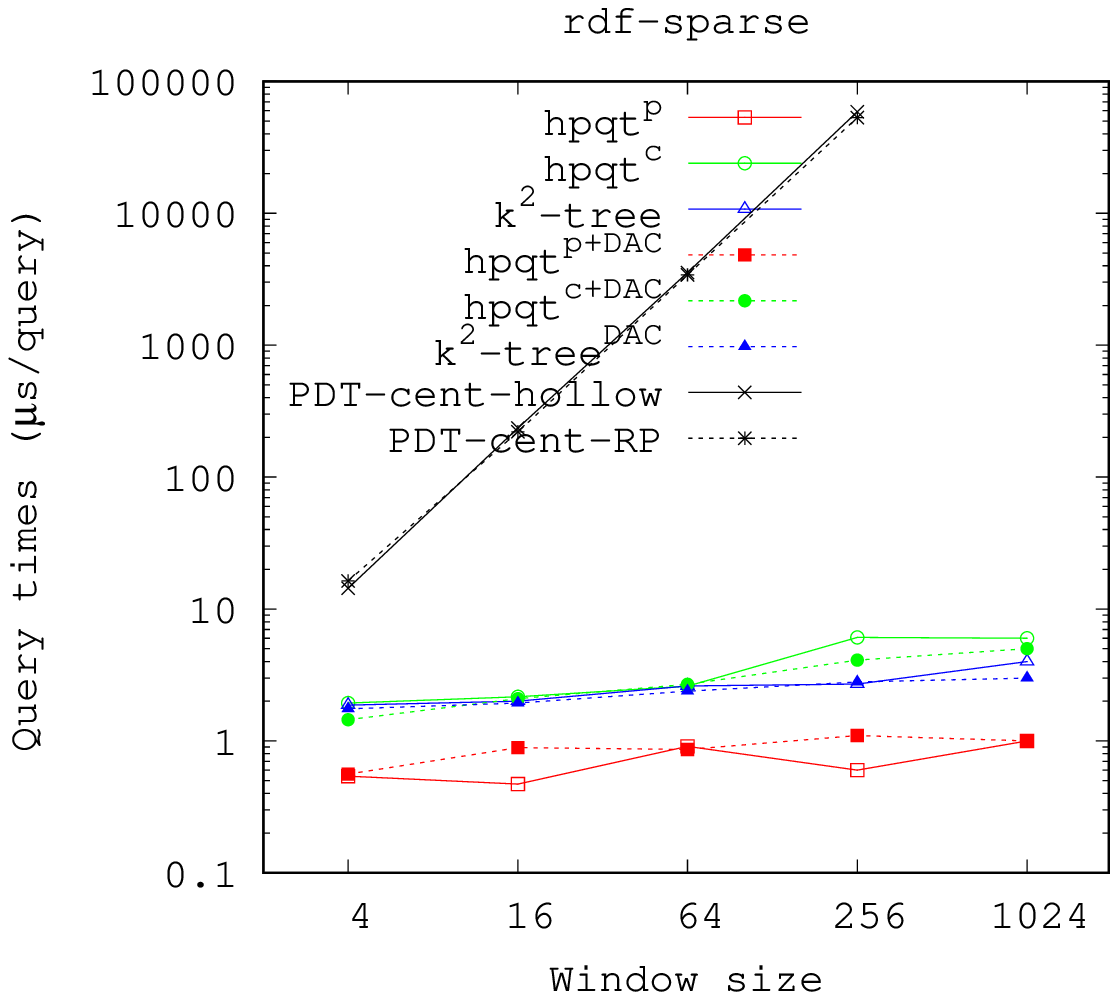}
    \includegraphics[angle=-90,width=0.49\textwidth]{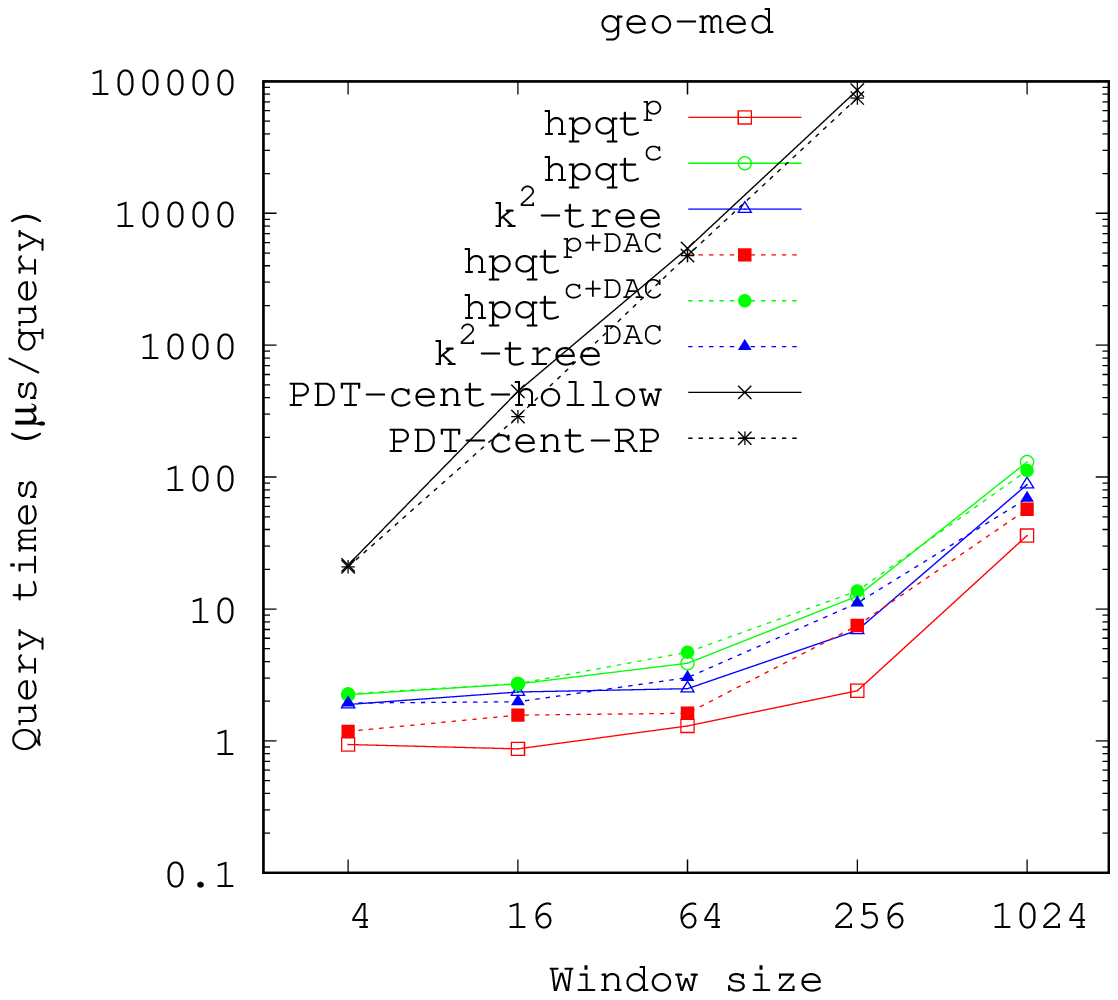}
    \includegraphics[angle=-90,width=0.49\textwidth]{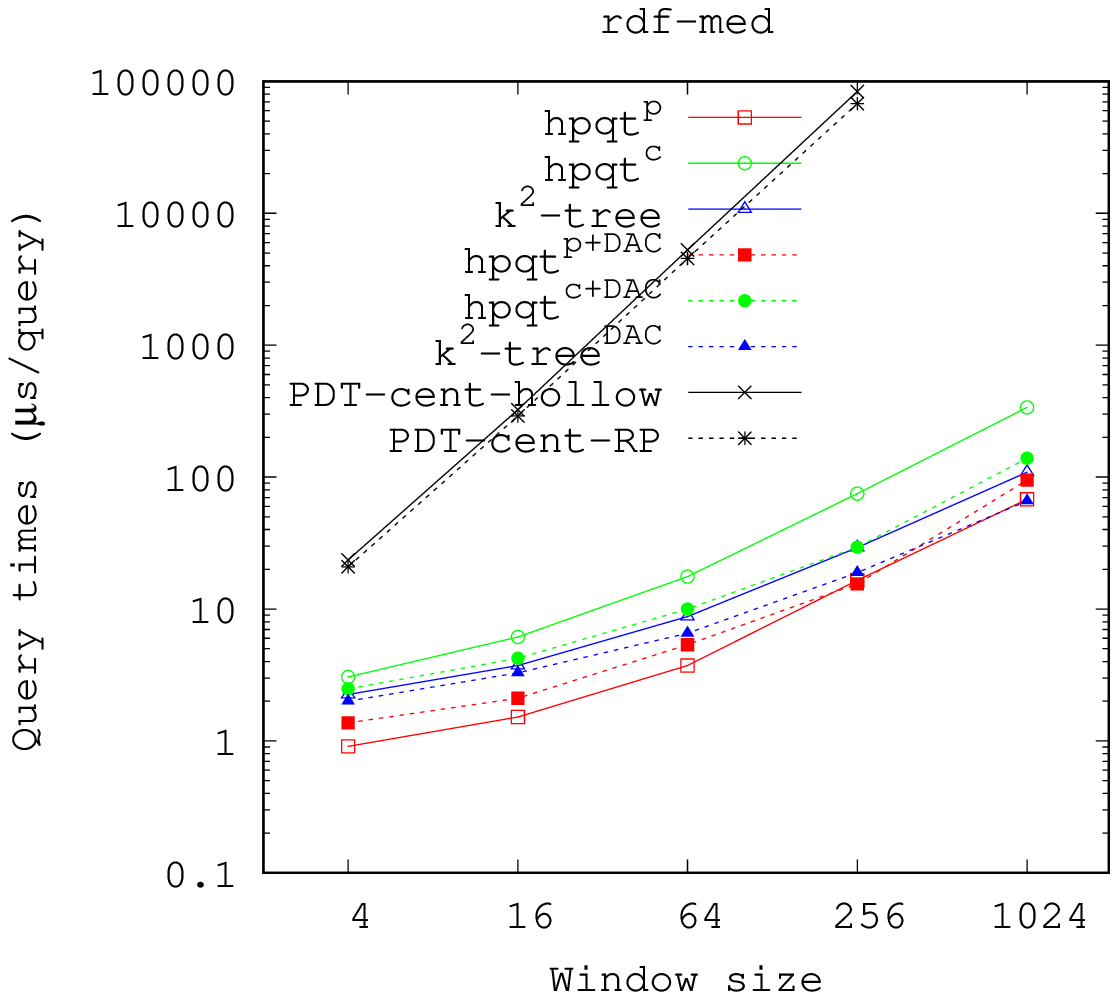}
    \includegraphics[angle=-90,width=0.49\textwidth]{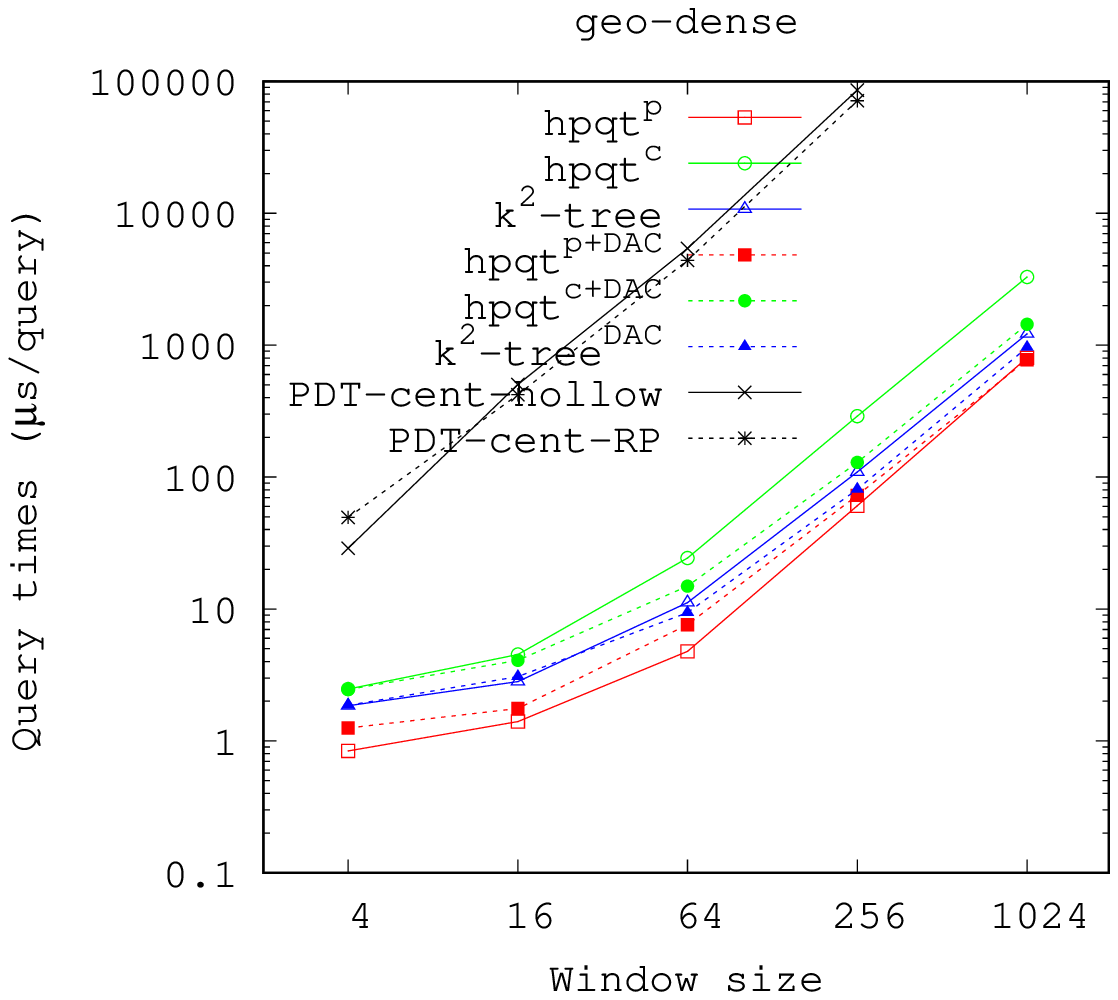}
    \includegraphics[angle=-90,width=0.49\textwidth]{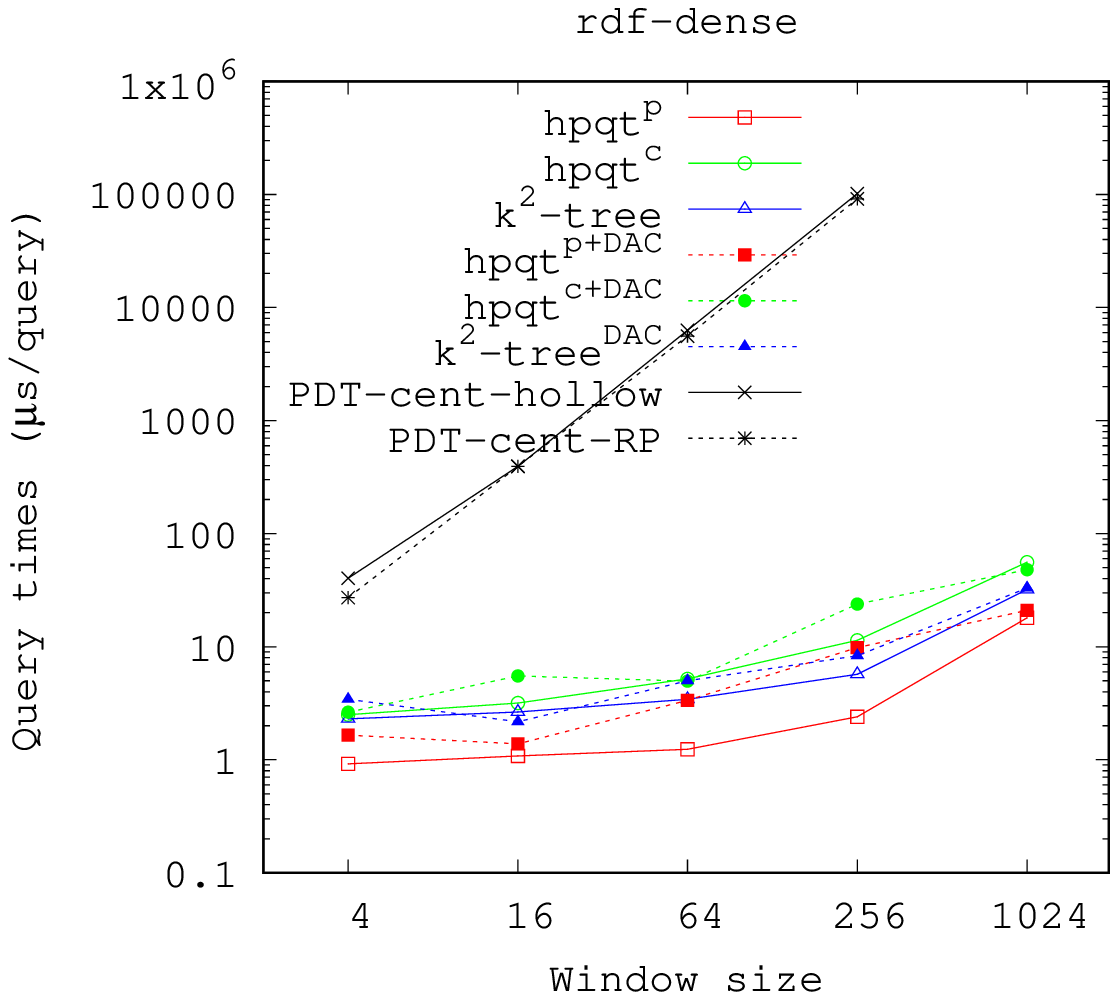}
  \caption{Query times for window queries, with varying window size, for GIS (left) and RDF (right) datasets. Results are in $\mu$s/query. Log-scale is used in both axis.}
  \label{fig:rangeGR}
\end{figure}

Figure~\ref{fig:rangeGR} displays the results on the GIS and RDF datasets, which are more difficult to compress. On those, \hpqtp and \hpqtpdac are the fastest in almost every case. The sparser datasets, displayed at the top, yield as expected the greatest difference in performance, with \hpqtp being 2--4 times faster than \ktdac. On the other hand, \hpqtR and \hpqtRdac are slower than \kt, but get very close. The \pdt variants are again much slower in all range queries.

We can conclude that the \hpqt is generally faster than the \kt at range queries, particularly for smaller windows. The \pdt structure is not competitive for these queries.

Other kinds of queries, such as row/column queries (requesting all points in a row/column of the grid), are frequent when representing graphs. On those queries the \ktree is slightly faster in general, since the efficiency of the \hpqt to locate the submatrix enclosing the query window does not produce any advantage. 

\subsection{Higher dimensions}

Finally, we test the applicability of our proposal to higher dimensions. We compare \hpqt with an implementation of the \koct, the extension of the \kt to 3 dimensions. We used a set of datasets, \mdta, \mdtb, and \mdtc, which had previously been evaluated for the \koct~\cite{BCBNP20}. Those 3-dimensional grids are obtained from elevation rasters, by considering the value stored in the raster of values as the third dimension. Table~\ref{tab:spaceraster} shows their main characteristics. 

\begin{table}[t]
\centering
\begin{tabular}{ l | r r r | r r r}
Dataset & Grid size ($r \times c \times d$) & Points \\
\hline
\mdta   & $4,001 \times 5,841 \times 578 $   &  23,051,888  \\
\mdtb   & $3,841 \times 5,841 \times 472 $   &  15,662,092 \\
\mdtc   & $7,721 \times 11,081 \times 978$   &  84,028,401  \\
\hline
\end{tabular}
\caption{Raster datasets used.}
\label{tab:spaceraster}
\end{table}

We will focus only on the variants including DAC compression (\hpqtpdac, \hpqtRdac, and \koctdac, the \koct with matrix vocabulary), because the \koctdac is the only available implementation of the \koct.

\begin{figure}[t]
 \centering
    \includegraphics[angle=-90,width=0.49\textwidth]{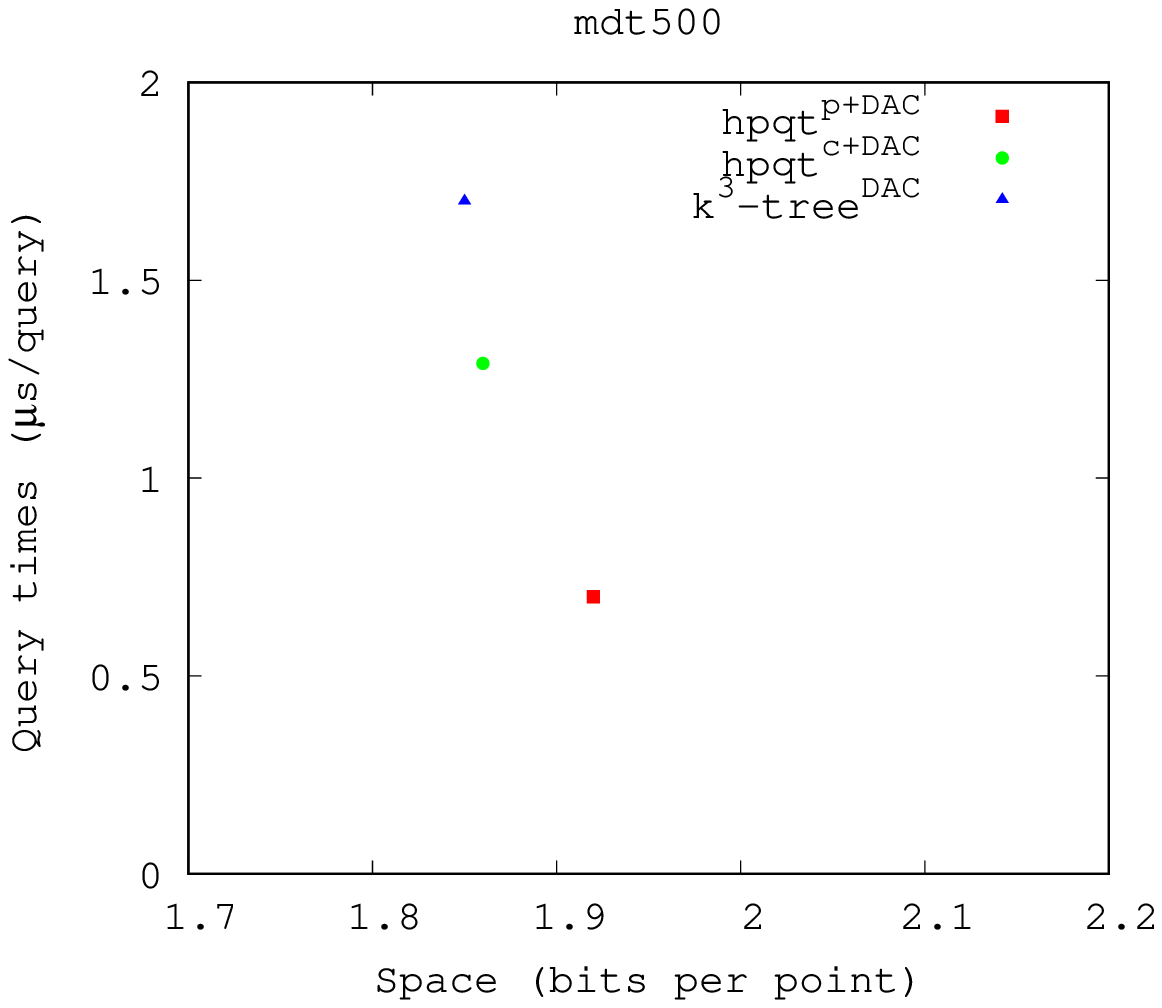}
    \includegraphics[angle=-90,width=0.49\textwidth]{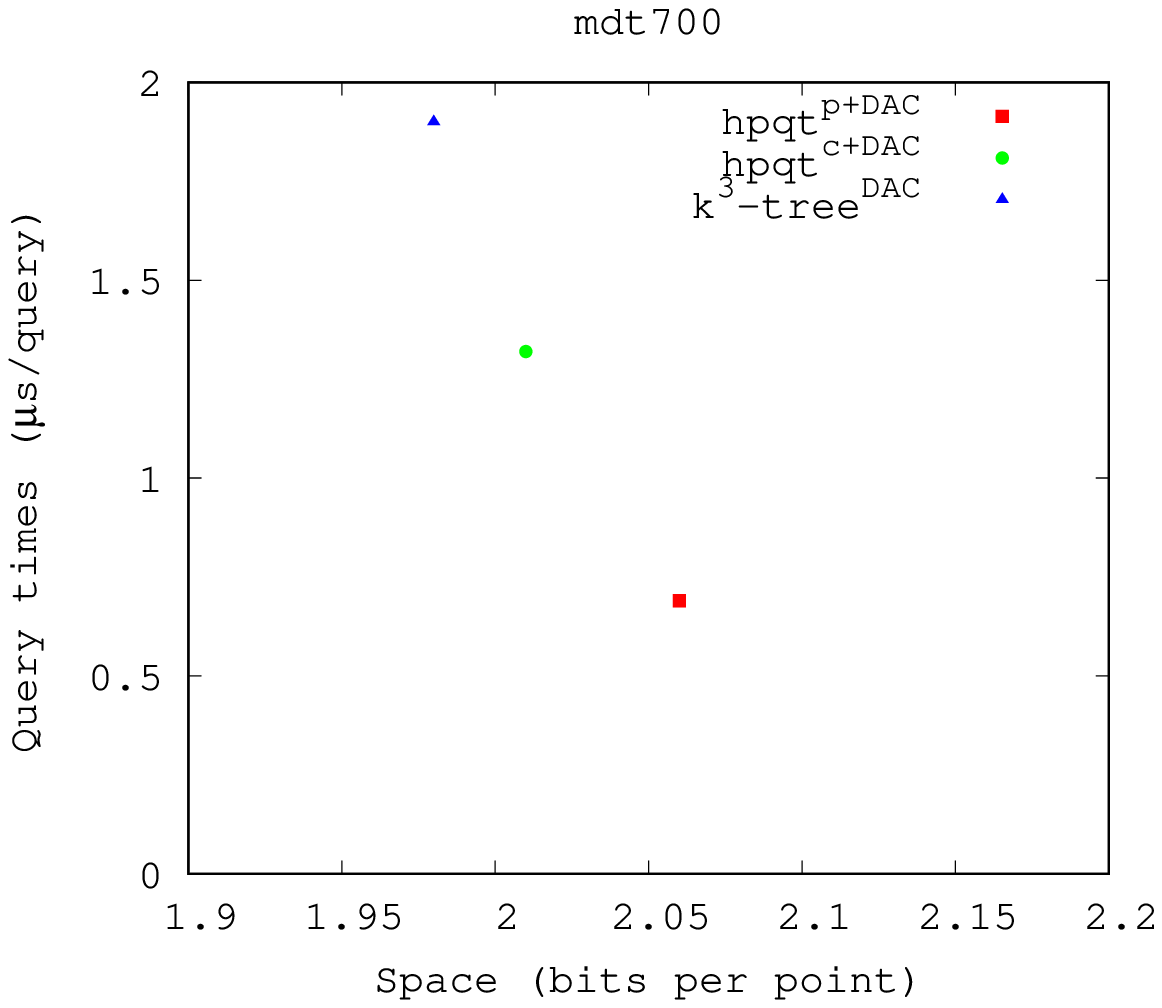}
     \includegraphics[angle=-90,width=0.49\textwidth]{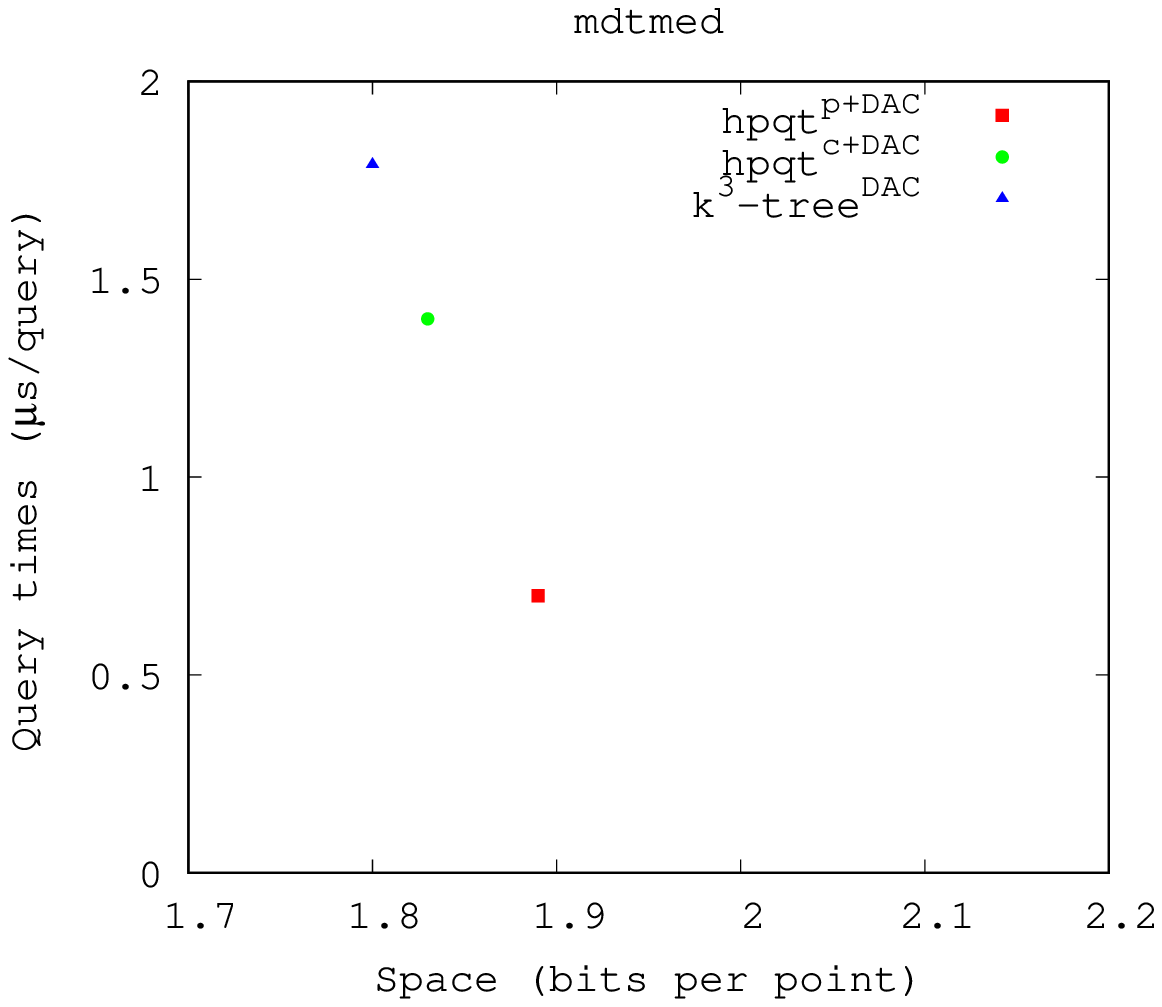}
  \caption{Space and times of membership queries for filled cells. Times are in $\mu$s/query.}
  \label{fig:rastermember}
\end{figure}

We first study compression and query performance for membership queries. For each dataset, we perform a membership query for each of the points it contains, and measure the average query time. Figure~\ref{fig:rastermember} displays the results obtained for all the datasets. The results are similar in all cases: \hpqtpdac and \hpqtRdac are slightly larger than \koctdac, but this difference is very small (less than 5\% for \hpqtpdac, and around 1\% for \hpqtRdac). On the other hand, both of our solutions are significantly faster than \koctdac: \hpqtpdac is about 2.5 times faster, and the compressed variant \hpqtRdac is still 25\% faster than \koctdac in all the datasets.

We now analyze the performance on range queries. For each dataset, we run sets of 100,000 random window queries, for different window sizes with the same side in all dimensions: $4 \times 4 \times 4$ to $256 \times 256 \times 256$. 

\begin{figure}[t]
 \centering
    \includegraphics[angle=-90,width=0.49\textwidth]{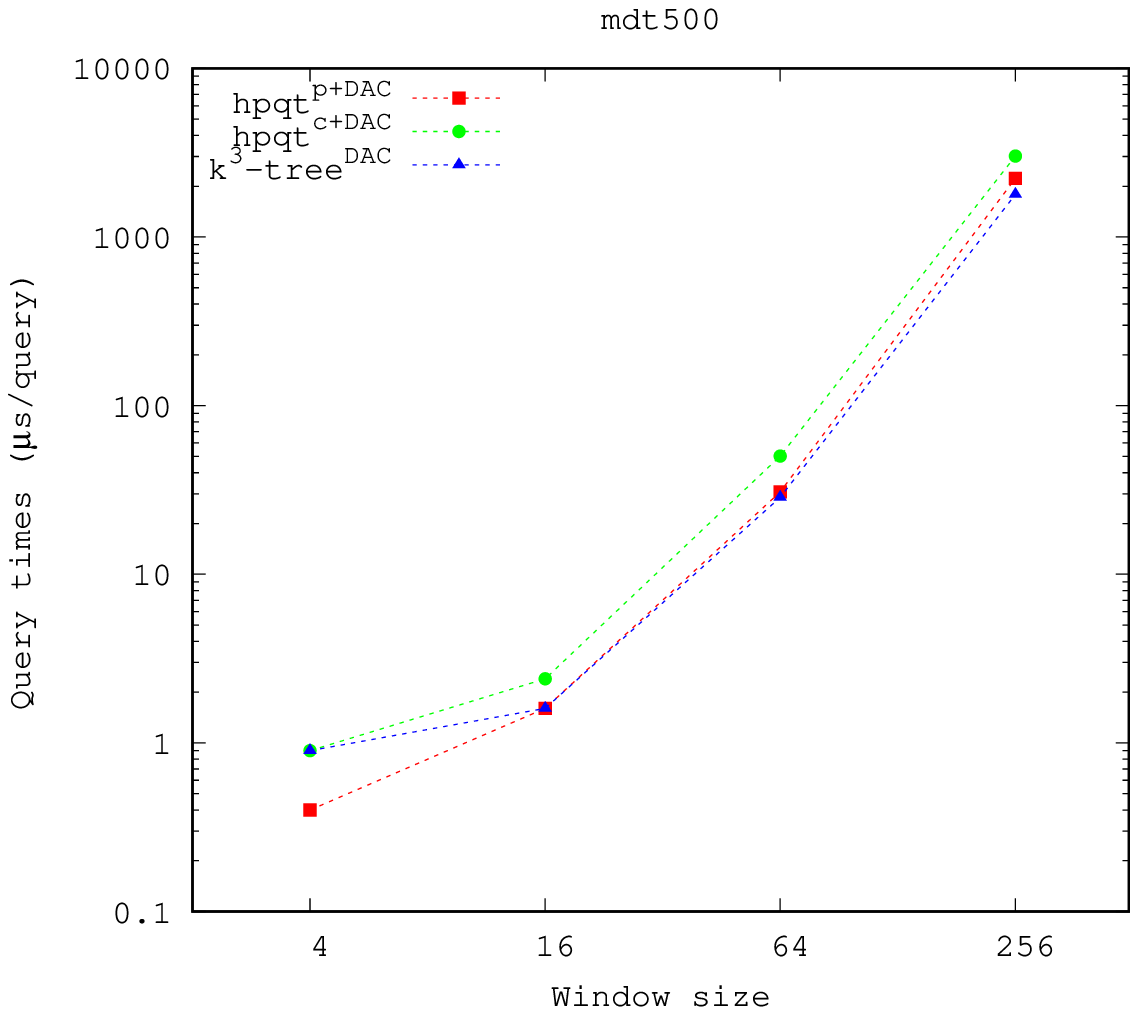}
    \includegraphics[angle=-90,width=0.49\textwidth]{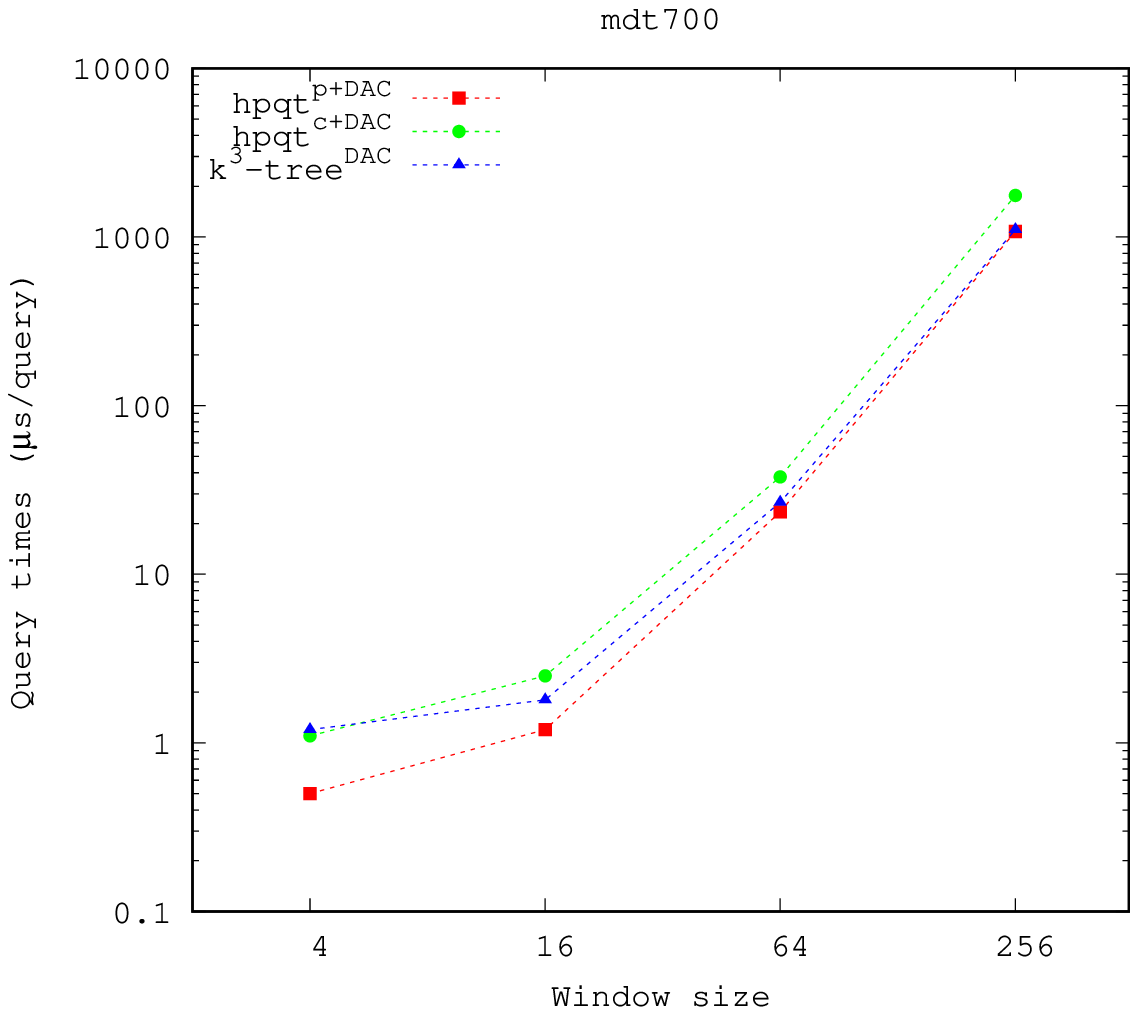}
    \includegraphics[angle=-90,width=0.49\textwidth]{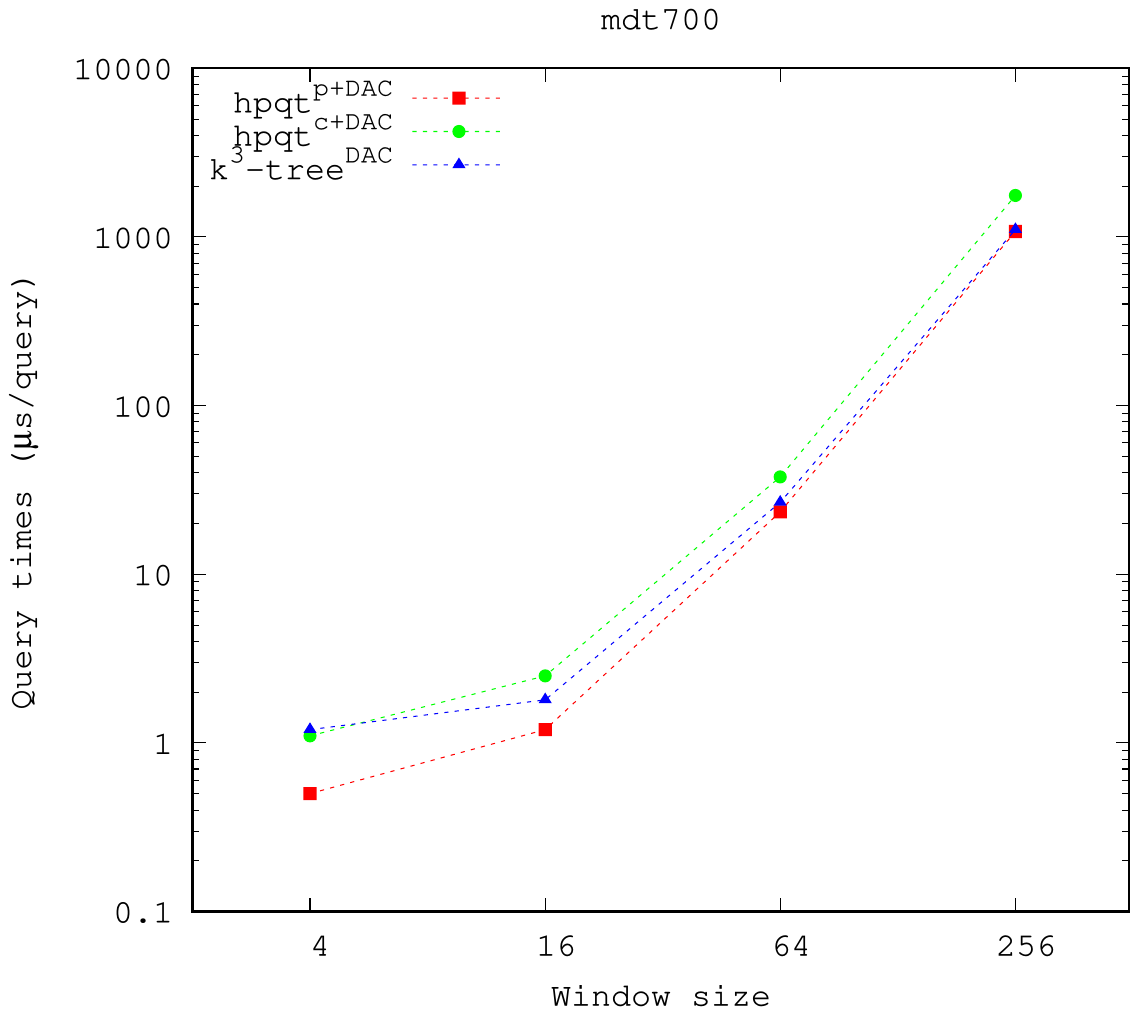}
  \caption{Times of raster queries for varying window sizes. Times are in $\mu$s/query. Log-scale is used in both axis.}
  \label{fig:rasterwindow}
\end{figure}

Figure~\ref{fig:rasterwindow} displays the query times on all the datasets, for varying window sizes. As on 2-dimensional data, \hpqt variants are faster for small query windows, and all the times become close on larger windows. In particular, \hpqtpdac is significantly faster than \koctdac in all cases for the smallest window sizes. The compressed variant, \hpqtRdac, is the slowest, but still close to \koctdac in all cases. 

Overall, we observe in general that the performance gap between \hpqt and \kt widens on three dimensions compared to the two-dimensional case, which is in line with our theoretical expectations.

\section{Conclusions}
\label{sec:conclusions}

We have introduced a fast space-efficient representation of quadtrees based on heavy-path decompositions, answering in the affirmative to the conjecture of Venkat and Mount~\cite{VM14}. Our structure represents a quadtree on $n$ points in a grid of size $u \times u$ using $\Oh{1}$ bits per quadtree node, and answers membership queries in $\Oh{\log n}$ time. Other compressed quadtree representations \cite{BLN14,VM14}, instead, require $\Oh{\log u}$ time, which can be significantly higher on sparse grids. We also prove that the space and time of our structure benefits from sparse and clustered point sets, which are common in various applications. Some, but not all, of those benefits extend to other quadtree representations as well.

We implemented our structure, demonstrating that it is also practical and competitive. The space requirements of our new representation are similar to other space-efficient representations of quadtrees, such as the $k^2$-trees \cite{BLN14}, but our structure is typically faster at retrieving existing points, especially isolated ones. Our structure is also generally faster to handle range queries, and on higher dimensions. Previous structures, instead, are faster when querying large empty areas of the grid.

One future work direction is to explore how the heavy path decomposition can be used to speed up other more sophisticated queries, like approximate-range and nearest-neighbor searches, by exploiting its ability to efficiently arrive at a desired submatrix.

Another interesting future work challenge is to make our structure dynamic, enabling point insertions and deletions, as done for the \kt and variants \cite{VM14,BCPdBN17,AdBGN19}. Every point insertion requires, in principle, marking that a new node has now two children (i.e., flipping a bit in some bitvector $L_d$) and adding a new path to the representation, somewhere inside bitvector $H$. Removing a point reverses this process. This can be supported in time $\Oh{\log u / \log\log u}$ if we use dynamic bitvectors \cite{NS12}, which is also the blowup factor induced on the other operations. Other approaches, which do not affect query times, might be possible \cite{CFRdBLN20}.

\section*{Acknowledgements}

Partially funded by the European Union’s Horizon 2020 research and innovation programme under the Marie Sk{\l}odowska-Curie grant agreement No 690941. GdB and SL funded by  MCIN/AEI/10.13039/501100011033 [grants PID2020-114635RB-I00 (EXTRACompact), PID2019-105221RB-C41 (MAGIST)], by MCIN/AEI/ 10.13039/501100011033, ``NextGenerationEU/PRTR'' [grants PDC2021-120917-C21 (SIGTRANS), PDC2021-121239-C31 (FLATCITY-POC)], by GAIN/Xunta de Galicia [grant ED431C 2017/53 (GRC)], and also supported by the Centro de Investigación de Galicia ``CITIC'', funded by Xunta de Galicia, FEDER Galicia 2014-2020 80\%, SXU 20\% [grant ED431G 2019/01 (CSI)].  TG funded by NSERC Discovery Grant RGPIN-07185-2020.  GN and DS funded by ANID -- Millennium Science Initiative Program -- Code ICN17\_002, Chile. GN funded by Fondecyt Grant 1-200038, Chile.


\begin{thebibliography}{10}
\expandafter\ifx\csname url\endcsname\relax
  \def\url#1{\texttt{#1}}\fi
\expandafter\ifx\csname urlprefix\endcsname\relax\def\urlprefix{URL }\fi
\expandafter\ifx\csname href\endcsname\relax
  \def\href#1#2{#2} \def\path#1{#1}\fi

\bibitem{Mor66}
G.~M. Morton, A computer oriented geodetic data base; and a new technique in
  file sequencing, Tech. rep., IBM Ltd. (1966).

\bibitem{Sam06}
H.~Samet, Foundations of Multidimensional and Metric Data Structures, Morgan
  Kaufmann, 2006.

\bibitem{Gar82}
I.~Gargantini, An effective way to represent quadtrees, Communications of the
  ACM 25 (1982) 905--910.

\bibitem{BLN14}
N.~Brisaboa, S.~Ladra, G.~Navarro, Compact representation of web graphs with
  extended functionality, Information Systems 39~(1) (2014) 152--174.

\bibitem{VM14}
P.~Venkat, D.~M. Mount, A succinct, dynamic data structure for proximity
  queries on point sets, in: Proc. 26th Canadian Conference on Computational
  Geometry (CCCG), 2014, p. article 32.

\bibitem{BCBNP20}
N.~R. Brisaboa, A.~Cerdeira-Pena, G.~de~Bernardo, G.~Navarro, O.~Pedreira,
  Extending general compact querieable representations to {GIS} applications,
  Information Sciences (2020) 196--216.

\bibitem{EGS08}
D.~Eppstein, M.~T. Goodrich, J.~Z. Sun, Skip quadtrees: Dynamic data structures
  for multidimensional point sets, International Journal of Computational
  Geometry and Applications 18~(1/2) (2008) 131--160.

\bibitem{GO14}
R.~Grossi, G.~Ottaviano, Fast compressed tries through path decompositions, ACM
  Journal of Experimental Algorithmics 19~(1).

\bibitem{Cla96}
D.~R. Clark, Compact {PAT} trees, Ph.D. thesis, University of Waterloo, Canada
  (1996).

\bibitem{Mun96}
J.~I. Munro, Tables, in: Proc. 16th Conference on Foundations of Software
  Technology and Theoretical Computer Science (FSTTCS), 1996, pp. 37--42.

\bibitem{WF90}
D.~S. Wise, J.~Franco, Costs of quadtree representation of nondense matrices,
  Journal of Parallel and Distributed Computing 9~(3) (1990) 282--296.

\bibitem{Kli71}
A.~Klinger, Patterns and search statistics, in: Optimizing Methods in
  Statistics, Academic Press, 1971, pp. 303--337.

\bibitem{HS79}
G.~M. Hunter, K.~Steiglitz, Operations on images using quad trees, IEEE
  Transactions on Pattern Analysis and Machine Intelligence 1~(2) (1979)
  145--153.

\bibitem{Nav16}
G.~Navarro, Compact Data Structures -- A practical approach, Cambridge
  University Press, 2016.

\bibitem{BLN13}
N.~Brisaboa, S.~Ladra, G.~Navarro, {DACs}: Bringing direct access to
  variable-length codes, Information Processing and Management 49~(1) (2013)
  392--404.

\bibitem{BDNR14}
N.~Bereczky, A.~Duch, K.~N{\'{e}}meth, S.~Roura, Quad-{K}-d trees, in: Proc.
  11th Latin American Symposium on Theoretical Informatics (LATIN), 2014, pp.
  743--754.

\bibitem{ST83}
D.~D. Sleator, R.~E. Tarjan, A data structure for dynamic trees, Journal of
  Computer and System Sciences 26~(3) (1983) 362--391.

\bibitem{FW94}
M.~L. Fredman, D.~E. Willard, Trans-dichotomous algorithms for minimum spanning
  trees and shortest paths, {Journal of Computer and System Sciences} 48~(3)
  (1994) 533--551.

\bibitem{PT06}
M.~P{\u{a}}tra\c{s}cu, M.~Thorup, Time-space trade-offs for predecessor search,
  in: Proc. 38th Annual ACM Symposium on Theory of Computing (STOC), 2006, pp.
  232--240.

\bibitem{OS07}
D.~Okanohara, K.~Sadakane, Practical entropy-compressed rank/select dictionary,
  in: Proc. 9th Workshop on Algorithm Engineering and Experiments (ALENEX),
  2007, pp. 60--70.

\bibitem{Knu09}
D.~E. Knuth, The Art of Computer Programming, volume 4: Fascicle 1: Bitwise
  Tricks \& Techniques; Binary Decision Diagrams, Addison-Wesley Professional,
  2009.

\bibitem{RRR}
R.~Raman, V.~Raman, S.~Rao, Succinct indexable dictionaries with applications
  to encoding \emph{k}-ary trees, prefix sums and multisets, {ACM} Transactions
  on Algorithms 3~(4) (2007) 43.

\bibitem{BoVWFI}
P.~Boldi, S.~Vigna, The {W}eb{G}raph framework {I}: {C}ompression techniques,
  in: Proc. 13th International World Wide Web Conference (WWW), 2004, pp.
  595--601.

\bibitem{BRSLLP}
P.~Boldi, M.~Rosa, M.~Santini, S.~Vigna, Layered label propagation: A
  multiresolution coordinate-free ordering for compressing social networks, in:
  Proc. 20th International Conference on World Wide Web (WWW), 2011, pp.
  587--596.

\bibitem{BCPdBN17}
N.~Brisaboa, A.~Cerdeira-Pena, G.~de~Bernardo, G.~Navarro, Compressed
  representation of dynamic binary relations with applications, Information
  Systems 69 (2017) 106--123.

\bibitem{AdBGN19}
D.~Arroyuelo, G.~de~Bernardo, T.~Gagie, G.~Navarro, Faster dynamic compressed
  $d$-ary relations, in: Proc. 26th International Symposium on String
  Processing and Information Retrieval (SPIRE), 2019, pp. 419--433.

\bibitem{NS12}
G.~Navarro, K.~Sadakane, Fully-functional static and dynamic succinct trees,
  ACM Transactions on Algorithms 10~(3) (2014) article 16.

\bibitem{CFRdBLN20}
M.~Coimbra, A.~Francisco, L.~Russo, G.~de~Bernardo, S.~Ladra, G.~Navarro, On
  dynamic succinct graph representations, in: Proc. 30th Data Compression
  Conference (DCC), 2020, pp. 213--222.

\end{thebibliography}
\end{document}